\documentclass[preprint,review,number,12pt]{elsarticle}
\usepackage{lineno}
\journal{NIM A}

\begin{document}
\begin{frontmatter}
	
\title{Solving inverse problems with the unfolding program TRUEE:\\Examples in astroparticle physics}
\author[tud]{N.~Milke\corref{cor1}}
\ead{natalie.milke@udo.edu}
\author[tud]{M.~Doert\corref{cor1}}
\ead{marlene.doert@udo.edu}
\author[ifae]{S.~Klepser}
\author[ifae]{D.~Mazin}
\author[uh]{V.~Blobel}
\author[tud]{W.~Rhode}
\address[tud]{Experimentelle Physik 5, Technische Universit{\"a}t Dortmund, 44221 Dortmund, Germany}
\address[ifae]{IFAE, Edifici Cn., Campus UAB, E-08193 Bellaterra, Spain}
\address[uh]{Institut f{\"u}r Experimentalphysik, Universit{\"a}t Hamburg, D-22761 Hamburg, Germany}
\cortext[cor1]{Corresponding author}

\begin{abstract}
The unfolding program TRUEE is a software package for the numerical solution of inverse problems. The algorithm was first applied in the FORTRAN\,77 program ${\cal RUN}$.
${\cal RUN}$ is an event-based unfolding algorithm which makes use of the Tikhonov regularization. It has been tested and compared to different unfolding applications and stood out with notably stable results and reliable error estimation. 
TRUEE is a conversion of ${\cal RUN}$ to C++, which works within the powerful ROOT framework. The program has been extended for more user-friendliness and delivers unfolding results which are identical to ${\cal RUN}$. Beside the simplicity of the installation of the software and the generation of graphics, there are new functions, which facilitate the choice of unfolding parameters and observables for the user.

In this paper, we introduce the new unfolding program and present its performance by applying it to two exemplary data sets from astroparticle physics, taken with the MAGIC telescopes and the IceCube neutrino detector, respectively. 

\end{abstract}


\end{frontmatter}

\linenumbers
\section*{Introduction}
Solving inverse problems can be described as a method to find the cause of known consequences. Problems of this kind manifest themselves in a wide range of research fields such as natural sciences, economics and engineering. Looking at physics as an exemplary field, inverse problems are among the fundamental challenges in various areas, for instance particle physics, crystallography or medicine. The particular problems and solutions in this paper will be presented and described alongside the subject of astroparticle physics. The nomenclature used here is mainly following~\cite{Blobel:1984p1896}.

The structure of this paper comprises three main sections. First, the class of inverse problems and the general procedure of unfolding with regularization are outlined. In a second section, the new unfolding program TRUEE is introduced. Subsequently, the first applications of the program in astroparticle physics, namely in the data analysis of the experiments MAGIC and IceCube, are presented in the third section. We conclude with a summary of the obtained results and an outlook on further extensions and applications of the program.

\section{Inverse problems and unfolding}
\label{sec:inverse}
In general, the distribution $f(x)$ of a variable $x$ has to be determined. However, it is often not possible to measure the value $x$ directly. Instead, the detector records $x$-correlated variables $y$. These signals can be seen as the mentioned consequences of the causation $x$. The goal is to get the best-possible estimate of the $f(x)$-distribution from the measured $g(y)$-distribution. As the measurement in a real experiment is distorted, this is not trivial. A direct allocation of a value $x$ to a value $y$ is not possible, because one $x$ value causes different signals with different $y$ values with certain probabilities. Furthermore, the probability to record a signal at all is usually less than one and depending on $x$, which causes a loss of events. Thus, the transformation of $x$ to $y$ is disturbed by a finite resolution and a limited acceptance of a real detector.

In mathematics, this problem can be described by the Fredholm integral equation \cite{fredholm}
\begin{eqnarray}
	\label{eq:fredholm}
g(y)=\int_c^dA(y,x)f(x)dx+b(y),	
\end{eqnarray}
where $g(y)$ is the distribution of the measured observable $y$ and can in general be multidimensional. The function $A(y,x)$ is called the kernel or response function and includes all effects which occur in a real measurement process. In most cases, this function is not known exactly and has to be determined by Monte Carlo (MC) simulations, where the measured and the real distributions are known. The parameters $c$ and $d$ are the integration limits of the range where $x$ is defined ($c\le x \le d$). The function $b(y)$ is the distribution of a possible background, which is assumed to be known.

In reality the measurement delivers discrete values. Furthermore the handling by the algorithm requires a numerical description of the distributions. Thus, a discretization of all functions is required. The distribution $f(x)$ can be parametrized with the Basis-spline (B-spline) functions $p_j(x)$ \cite{deBoor:2001p1990} and the corresponding coefficients $a_j$

\begin{eqnarray}
	\label{eq:b-splines}
	f(x)=\sum_{j=1}^ma_jp_j(x).
\end{eqnarray}
The B-spline functions consist of several polynomials of a low degree. In the following cubic B-splines are used. They consist of four polynomials of third degree each. The points where adjacent polynomials overlap are called knots. At the knot positions a B-spline is continuously differentiable up to the second derivative, which is important because the second derivative is used for the implemented regularization (see Eq.~\ref{eq:curvature}). For equidistant knots, the cubic B-splines are bell-shaped. Because of the low degree of the polynomials, an interpolation with B-spline functions does not tend to oscillate. By using this parametrization, the B-spline functions can be included in the response function during the discretization:

\begin{eqnarray}
	\label{eq:A-parametrized}
	\int_c^dA(y,x)f(x)dx&=&\sum_{j=1}^ma_j\left[\int_c^dA(y,x)p_j(x)dx\right] \nonumber
	\\&=&\sum_{j=1}^ma_jA_j(y).
\end{eqnarray}
By integrating over the $y$-intervals, the kernel function becomes a response matrix:  
\begin{eqnarray}
	\label{eq:A-element}
	A_{ij}=\int_{y_{i-1}}^{y_i}A_j(y)dy.
\end{eqnarray}
The same integration can be carried out for the measured distribution $g(y)$ and the background distribution $b(y)$:
\begin{eqnarray}
	g_i=\int_{y_{i-1}}^{y_i}g(y)dy,\\
	b_i=\int_{y_{i-1}}^{y_i}b(y)dy.
\end{eqnarray}
Consequentially, the Fredholm integral equation becomes the matrix equation
\begin{equation}
	\label{eq:matrix}
	\bf{g}=\bf{Aa}+b,
\end{equation}
with $\bf{g}$, $\bf{a}$ and $\bf{b}$ as vectors and $\bf{A}$ as the response matrix. To determine the sought distribution $f(x)$, the coefficients $a_j$ need to be found.

Solving Eq.~\ref{eq:matrix} is called unfolding and is generally not trivial. Due to the finite resolution a smoothing effect on the measured distribution $\bf{g}$ is introduced. After the rearrangement of the matrix equation this smoothing effect is inverted and results in implausible oscillations of the sought distribution $f(x)$. The most straightforward approach for the solution is the inversion of the response matrix $\bf{A}$, if $\bf{A}$ is quadratic and non-singular. The resulting inverse matrix $\bf{A^{-1}}$ contains negative non-diagonal elements and very large diagonal elements. This causes the mentioned oscillation, which appear in any approach of solving Eq.~\ref{eq:matrix} if no additional corrections are applied.
This is known as a so-called ill-posed problem and generally occurs in all measurement processes. 

To suppress the oscillations in the unfolded distribution, so-called regularization methods are applied. In the presented realization the Tikhonov regularization \cite{tikhonov1977solutions} is implemented. The method, in its generalized form, requires the linear combination of the unfolding term with a regularization term (sometimes called penalty term), which contains a regularization factor. The regularization term contains an operator, which implies some a-priori assumptions about the solution, such as smoothness. In the current case the smoothness of the solution is controlled by the curvature operator $\bf{C}$. A large curvature corresponds to large oscillations. Thus, reduction of curvature implies reduction of oscillations and that smoothes the resulting distribution. Since the parametrization of $f(x)$ is based on cubic B-spline functions, the curvature $r(\bf{a})$ takes the simple form of a matrix equation
\begin{equation}
	\label{eq:curvature}
	r({\bf{a}})=\int \left( \frac{d^2f(x)}{dx^2} \right)^2dx={\bf{a}}^T{\bf{Ca}},
\end{equation}
with $\bf{C}$ as a known, symmetric, positive-semidefinite curvature matrix.

The actual unfolding is performed as follows. At first the response matrix $\bf{A}$ is calculated, based on the MC sample. To determine the coefficients $\bf{a}$ of the final result, the unfolding equation (Eq.~\ref{eq:matrix}) is set up, where $\bf{g}$ is the real measured observable distribution. To fit the right hand side to the left hand side of this equation, a maximum likelihood fit is performed. For simplicity, a negative log-likelihood function 
\begin{equation}
	\label{eq:log-likelihood}
	S({\bf{a}})=\sum_i(g_i({\bf{a}})-g_{i,m}\,\ln g_i({\bf{a}}))
\end{equation}
is formed and minimized. Here $g_{i,m}$ is the number of measured events in an interval $i$ including the possible background contribution in this region. This number follows the Poisson distribution with mean value $g_i$. A Taylor expansion of the negative log-likelihood function can be written as
\begin{eqnarray}
	\label{eq:taylor-likelihood}
	S({\bf{a}})&=&S({\bf{\tilde a}})+({\bf{a}-\tilde a})^T{\bf{h}} \nonumber
	\\&+&\frac{1}{2}({\bf{a}-\bf{\tilde a}})^T{\bf{H}(\bf{a}-\bf{\tilde a}}) + ..,
\end{eqnarray}
with gradient $\bf{h}$, Hessian matrix $\bf{H}$ and $\bf{\tilde a}$ as a first estimation of coefficients, which have to be found.

After considering regularization (Eq.~\ref{eq:curvature}), the final fit function 
\begin{eqnarray}
	\label{eq:unfoldingfit}
	R({\bf{a}})&=&S({\bf{\tilde a}})+({\bf{a}-\tilde a})^T{\bf{h}}+\frac{1}{2}({\bf{a}-\tilde a})^T{\bf{H}}({\bf{a}-\tilde a})\nonumber
	\\&+&\frac{1}{2}\tau {\bf{a}}^T{\bf{Ca}}
\end{eqnarray}
has to be minimized to obtain the unfolded result. The regularization parameter $\tau$ controls the effect of the regularization. The challenge is to find a proper value for $\tau$, to get an optimal estimation of the result as a balance between oscillations and the smoothing effect of the regularization.

One method to define a value for $\tau$ is to set up the relation between $\tau$ and the effective number of degrees of freedom $ndf$
\begin{equation}
	\label{eq:tau_ndf}
	ndf = \sum_{j=1}^{m} \frac{1}{1+\tau S_{jj}}.
\end{equation}
Here $S_{jj}$ are the eigenvalues of the diagonalized curvature matrix $\bf{C}$, arranged in increasing order. The summands in Eq.~\ref{eq:tau_ndf} can be considered as filter factors for the coefficients. These coefficients represent the transformed measurement and are arranged in decreasing order. The filter factors with values $< 1$ diminish the influence of insignificant coefficients. Accordingly an increasing value of $\tau$ cuts away smoothly the high order coefficients and reduces the number of degrees of freedom. In turn, the definition of number of degrees of freedom allows the specification of the number of filter factors and thus of the regularization strength.

To obtain Eq.~\ref{eq:tau_ndf}, the Hesse and curvature matrices in Eq.~\ref{eq:unfoldingfit} have to be diagonalized simultaneously. To do this, a common transformation matrix has to be found, which transforms the Hesse matrix into a unit matrix and diagonalizes the curvature matrix \cite{Blobel:1998p1987}.

A lower limit of the parameter $\tau$ can be estimated by testing the statistical relevance of the eigenvalues of the response matrix. Applying Eq.~\ref{eq:tau_ndf}, the number of degrees of freedom has to be chosen such that $\tau$ is above the suggested limit, in order to avoid the suppression of significant components in the solution.

\section{TRUEE}
\label{Sec:truee}
Several algorithms have been developed for solving inverse problems in different categories. One of them is ${\cal RUN}$ - {\bf{R}}egularized {\bf{UN}}folding \cite{Blobel:1984p1896} \cite{Blobel:1998p1987}, which uses the mathematics outlined in Sec.~\ref{sec:inverse}. ${\cal RUN}$ was developed in the 1980's in FORTRAN\,77 and was updated several times, the last time around 1995~\cite{Blobel:1996p1382}. The ${\cal RUN}$ algorithm has been converted to a ROOT based C++ version. Furthermore it has been equipped with additional functions and user-friendly extensions. This new software package is called TRUEE - {\bf{T}}ime-dependent {\bf{R}}egularized {\bf{U}}nfolding for {\bf{E}}conomics and {\bf{E}}ngineering problems.

The algorithm can process event-wise data input, by reading single n-tuples of variables. This flexibility permits an individual determination of the response matrix for every specific case, in contrast to algorithms which can only deal with histograms as input. Additionally, supplementary cuts or event weights can be applied internally without changing the input data files. Furthermore, the availability of individual event information allowed the development of a method to verify the unfolding result (see Sec.~\ref{Sec:verification}). A set of up to three observables can be used to perform the unfolding fit. Thus, the precision of the estimated function can be enhanced by choosing three observables with complementary information content.

\begin{figure}[!thh]
	\centerline{\includegraphics[width=3.in]{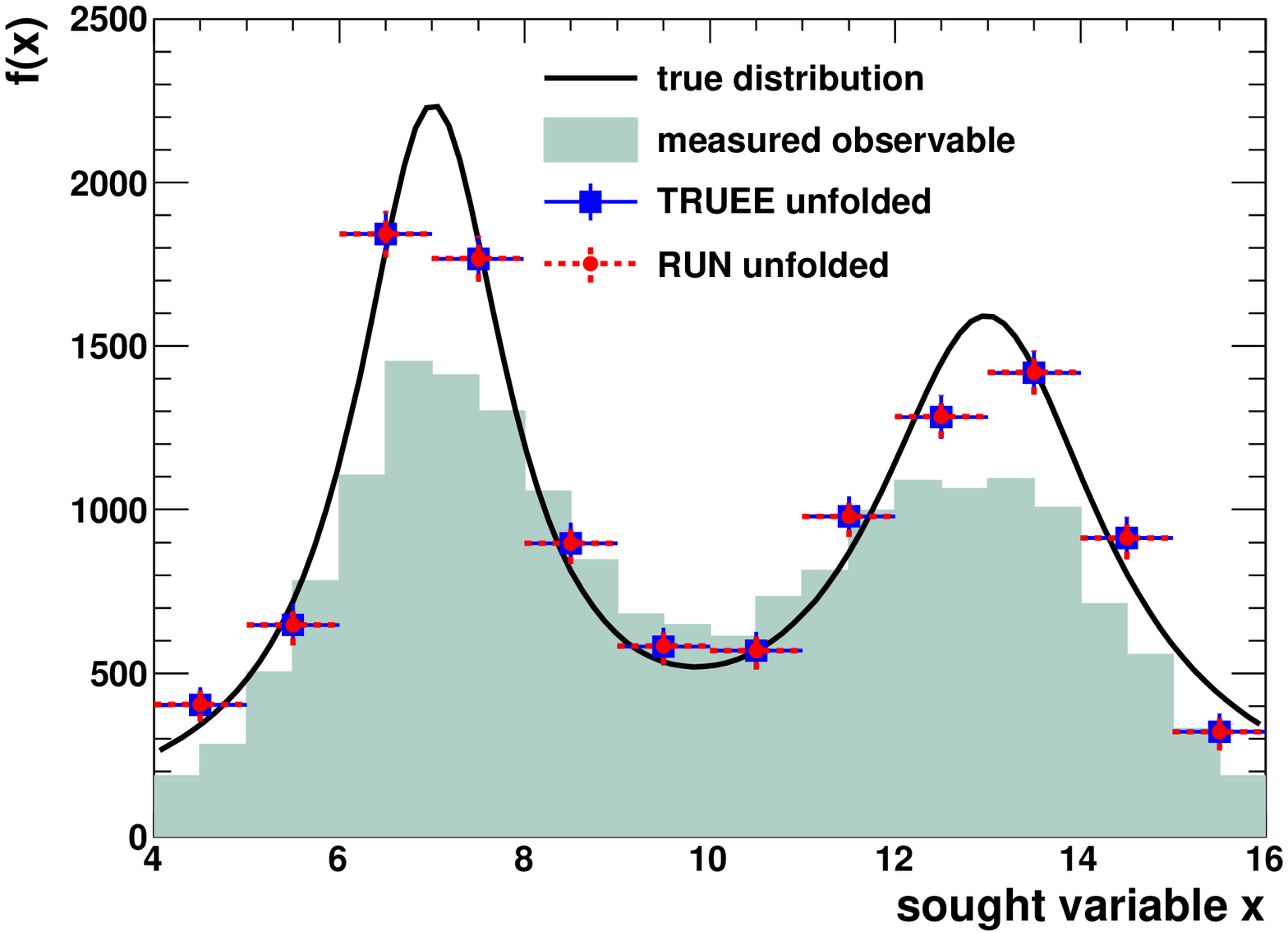}}
\vspace{1pt}
	\centerline{\includegraphics[width=3.in]{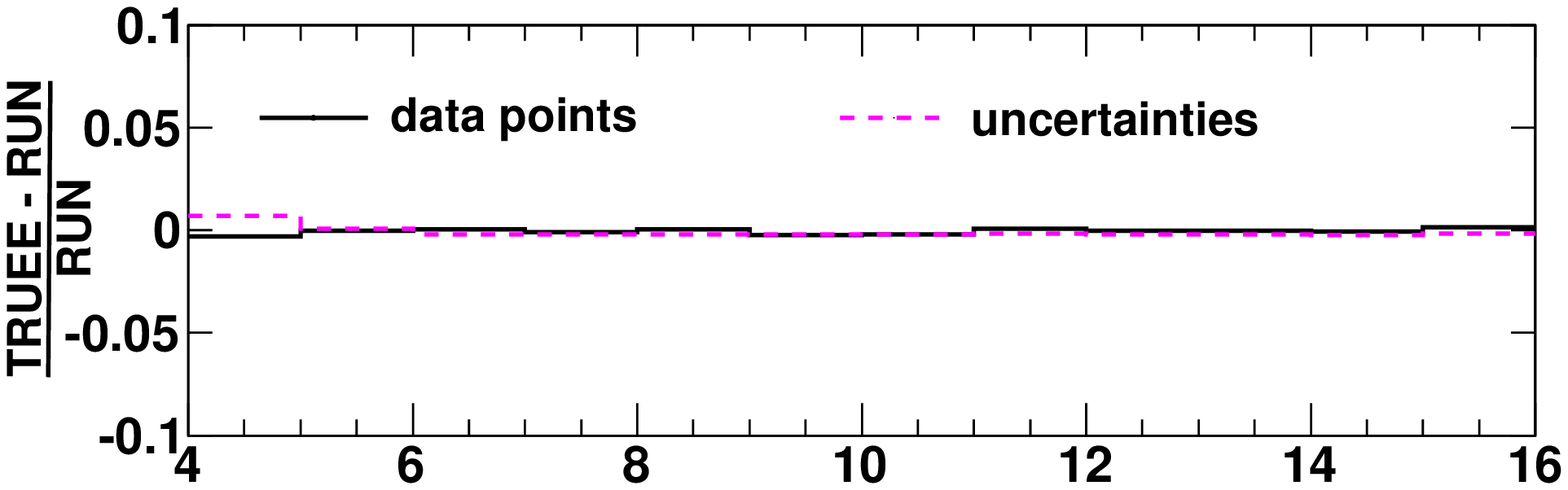}}
	\caption{Comparison of results of the original unfolding algorithm ${\cal RUN}$ (gray circles) and the new C++ version TRUEE (black squares). The solid line shows the true sought distribution. The shaded area represents the distribution of the measured observable, which has been used for the unfolding. The relative deviations of the bin contents and uncertainties from both algorithms can be seen in the lower figure and show a good agreement between the unfolding results.}
	\label{Fig:comparison}
\end{figure}

${\cal RUN}$ and TRUEE deliver the same results in terms of both data points and uncertainty estimation. In Fig.~\ref{Fig:comparison} the comparison of the two algorithms is demonstrated by performing an unfolding of a simu\-la\-ted distribution. The true distribution and the observable are shown as well, illustrating the finite resolution and limited acceptance of the simulated measurement. In the bottom panel, the ratios of the unfolded bin contents show an almost perfect matching between the algorithms. Minor deviations can be accounted to the distinct handling of floating point variables of the two different compiler types.

Instead of unfolding a given data sample as a whole, TRUEE can also be used to investigate changes with time in the investigated distribution. If structural interruptions with respect to time are found beforehand, TRUEE can unfold time slices of the inspected data and reveal time-correlated changes in the corresponding distribution. 

The installation of TRUEE is straight forward on UNIX based operating systems, as it uses CMake \cite{Martin:2003p1633}. 
The new algorithm is able to deal with two different types of input files: ASCII and ROOT files.
To make the analysis procedure more comfortable, new functions have been included, which are described in the following (Sec.~\ref{Sec:obs-selection} to \ref{Sec:testmode}).
Besides the newly implemented functionalities, a well-proven ${\cal RUN}$ function for the verification of the unfolding result shall also be mentioned. The functionalities of a built-in correction for the acceptance of the experimental setup and the treatment of background that is present in the measurement are likewise inherited from ${\cal RUN}$ and will be described below as well.

\subsection{Selection of observables}
\label{Sec:obs-selection}
Generally a measured event is characterized by a large set of observables. TRUEE can deal with more than~30 different observables, of which up to three can be used for the unfolding at the same time. These should be the observables which are most correlated with the variable to be unfolded. To check the dependency of the observables on this variable, correlation and profile histograms are automatically created from the MC sample. Different examples of such histograms are shown in section~\ref{Sec:experiments}.

\subsection{Parameter selection}
\label{Sec:parameter_selection}
Generally, an unfolding algorithm requires the input of various parameters by the user, such as the binning of observables and final histogram as well as the influence of the regularization. The challenge of selecting an optimal pa\-ra\-me\-ter set has its difficulties in finding a result with low correlation between the unfolded data points and low bias, which is introduced by re\-gu\-la\-ri\-za\-tion. This outcome has to be identified out of many results with different parameter combinations. The three crucial parameters are
\begin{itemize}
	\item number of bins
	\item number of knots
	\item number of degrees of freedom.
\end{itemize}
The number of knots defines the number of B-splines used in the superposition for the unfolded function (see Eq.~\ref{eq:b-splines}). This number is related to the internal binning of the sought distribution $f(x)$ for the unfolding, which is chosen to be equidistant. After the estimation, the obtained $f(x)$ is transformed to a binned distribution that represents the final result, for which the number of bins can be chosen. The individual bins can have different widths. 

The number of degrees of freedom controls the influence of the regularization by defining the parameter $\tau$ (see Eq.~\ref{eq:tau_ndf}). A low number of degrees of freedom means strong regularization and a positive correlation between the unfolded data points. Thus, the introduced bias may be too high. A large number reduces the regularization and causes implausibly large fluctuations and uncertainties. Oftentimes, these can even be larger than the bin contents due to negative correlations between data points. A balance between these extreme cases has to be found. In general the number of degrees of freedom should be roughly the same as the number of bins, as this ensures that on the one hand no information which is contained in the measurement is discarded and on the other hand no positive correlations are introduced by solving an underdetermined system.

\begin{figure}[!thh]
	\centerline{\includegraphics[width=3.in]{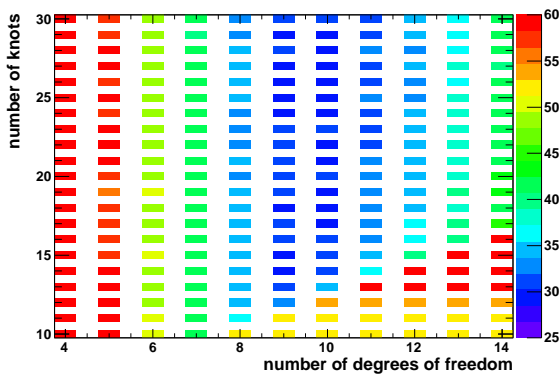}}
	\caption{The correlation-related value $\kappa$, as a quality factor for the unfolding result, color-coded in the two-dimensional histogram of varying number of knots and number of degrees of freedom. In this example, the best results with the lowest correlations are located in the range between 9 and 11 degrees of freedom. The strong dependency of correlation on the regularization, expressed by the number of degrees of freedom, is clearly visible.}
	\label{Fig:checkPara}
\end{figure}

To facilitate the task of choosing a good combination in the three parameters outlined above, histograms are provided by TRUEE, which show a quality value $\kappa$ that indicates whether the correlations among the unfolded data points can be neglected.
For each number of bins, one such histogram is provided, where the number of knots and the number of degrees of freedom are the two dimensions of the histogram. In these histograms, the parameter region, where the least correlation between the data points can be seen, can be identified. An example of such a chart is shown in Fig.~\ref{Fig:checkPara}.

The displayed correlation-related value $\kappa$ is the resulting quantity of a test to determine whether the covariance matrix can be considered as diagonal, which has been developed within the original algorithm of ${\cal RUN}$. 
Within the test, 5\,000 multivariate gaussian deviations of the unfolded result are randomly generated using the full covariance matrix. Each of the multivariate deviations is compared to the unfolded result by a $\chi^2$ calculation, in which the covariance matrix is assumed to be diagonal. The p-values obtained from the $\chi^2$ values are filled into a histogram. In the case of a flat distribution of p-values the covariance matrix can be considered as diagonal and the correlations between the unfolded data points are negligible. The value $\kappa$ describes the flatness of the p-value histogram. It is built as the sum of the absolute residuals between the determined and a flat p-value distribution divided by the number of bins of the p-value histogram.

\subsection{Test mode}
\label{Sec:testmode}
\begin{figure}[!thh]
	\centerline{\includegraphics[width=3.in]{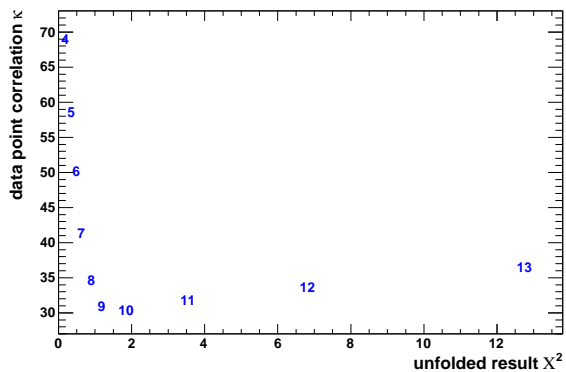}}
	\caption{The correlation-related value $\kappa$ versus the $\chi^2$ value from the comparison of the unfolded result with the true distribution. The bin contents are marked with the number of degrees of freedom. The optimal result can be found in the region of small $\kappa$ and small $\chi^2$ values, here 9 degrees of freedom. The additional variation of numbers of knots and numbers of bins is not shown in this figure.}
	\label{Fig:kappaVSchi}
\end{figure}

To find an optimal set of unfolding parameters and check whether the unfolding is working well with the selected observables, a test mode has been implemented as an additional tool. While running in test mode, the simulated event sample, which is given to the program, is considered alone. In unfolding mode, this sample is only used to determine the detector response matrix. In test mode, a fraction of events from the simulated sample can be selected to serve as a pseudo data sample, which is subsequently unfolded. The rest of the simulated events is used to determine the response matrix in the usual way. Since the true distribution of the variable which is to be unfolded is known in a simulation, it is possible to test whether the unfolded distribution matches the true one. The comparison between the unfolded and the true distributions is performed with a Kolmogorov-Smirnov test~\cite{Chakravarti:1967} and a $\chi^2$ test. Histograms showing the agreement of the distributions for each combination of number of knots and degrees of freedom are provided. In test mode an additional parameter selection method can be used by plotting $\kappa$ versus the values of the $\chi^2$ test for each parameter set. The parameter sets with minimal correlation among the data points and a good fit can be found where both the $\kappa$ and the $\chi^2$ value are small. An example is given in Fig.~\ref{Fig:kappaVSchi}. The parameter setting, which shows the best agreement in the test unfolding, should be used for the actual unfolding of the real data.

\subsection{Verification}
\label{Sec:verification}
Generally, the distributions in detector observables of the MC do not necessarily match the ones in measured data.
After the unfolding result has been obtained, the consistency of the unfolding and the MC simulation can be verified by comparing the distributions of individual observables of real data with a weighted MC sample. To do this, the MC sample, which has been used to calculate the response matrix, is weighted with respect to the distribution in the variable $x$ that is seen in the unfolding result. Hence, following the same distribution in $x$, the resulting MC sample should describe the data sample perfectly well and all observable distributions should match between MC and data. This is especially interesting for observables that have not been considered during the unfolding fit. Histograms showing distributions of real data and the weighted MC sample are provided for each observable that has been introduced to the program. Examples are shown in Sec.~\ref{sec:verificationMagic} and~\ref{sec:verificationIcecube}.

\subsection{Acceptance Correction}
\label{Subsec:acceptance}
For the reconstruction of the initial distribution $f(x)$, which describes the sought physically meaningful quantity, it is necessary to consider the limited acceptance and loss of events due to a quality selection during the analysis. A corresponding correction can be done by TRUEE, if the function of the generated MC event distribution is supplied by the user. 
The acceptance of the measurement, which can be a function of the variable to be unfolded, is defined as the ratio between the generated MC event distribution and the MC event distribution at the final analysis level. TRUEE determines this acceptance for each bin of the demanded variable and applies it during the unfolding of the distribution.

\subsection{Consideration of background}
If background is present in the measurement process, it has to be taken into account during the unfolding. With a given background event sample, TRUEE performs a corresponding correction. By adding the detector observable distributions of the background sample to the expectation (see Eq.~\ref{eq:matrix}), it is considered during the unfolding fit.

\section{Application of TRUEE in astroparticle physics experiments}
\label{Sec:experiments}
Many ground based astroparticle physics detectors suffer from the fact that it is not possible to directly measure the primary particles and their properties. Indirect detection methods are necessary, which instead utilize atmospheric particle showers or measurements of secondary particles. The correlation between the distributions in the thus derived observables and the distribution in the sought quantities is usually complex and ambiguous. Moreover, detection processes are affected by limited acceptance. For example, the original particle's energy and direction are folded with the interaction cross sections and response of the detector. Thus, the application of unfolding methods is necessary to determine the distribution of the variable to be found. In this section we present the application of TRUEE in the astroparticle experiments MAGIC \cite{performancepaper} and IceCube \cite{Halzen:2010p1795}.

\subsection{The MAGIC telescopes}
\label{sec:magic}

\subsubsection{The experiment}
\label{subsubsec:magicexp}
The \textbf{M}ajor \textbf{A}tmospheric \textbf{G}amma-ray \textbf{I}maging \textbf{C}herenkov telescopes are a stereoscopic system of two Cherenkov telescopes, which is situated on the Roque de los Muchachos on the Canary island of La Palma. MAGIC started its operation as a single telescope experiment in 2004 and has been upgraded to a stereoscopic system later on, which is operational since late 2009.

The experiment accesses the energy range of 50\,GeV to several tens of TeV in cosmic gamma-rays in the standard operation mode. Measurements of the gamma-ray flux at these energies give insight into a large set of highly energetic astronomical sources, such as Supernova Remnants, Active Galactic Nuclei and potentially Gamma-Ray Bursts. Besides source studies, gamma-rays also allow the investigation of the extragalactic background light and more exotic phenomena like the search for dark matter particles.

Ground-based gamma-ray detectors like MAGIC exploit the Earth's atmosphere as their detection volume. High energy gamma-rays reaching Earth cause atmospheric particle showers. These are accompanied by Cherenkov radiation, which can be detected by the telescope cameras. 
The number of Cherenkov photons, along with the reconstruction of the shower geometry, can deliver an energy estimation of each incident gamma-ray.

Unfortunately, these events are outnumbered by a huge background of hadronically induced particle showers, which have to be separated statistically from the sought gamma particle showers in the course of the so-called gamma/hadron separation \cite{Bock:multivariate},\cite{Voigt:cut-own},\cite{magic:rf}. 

Additional background from diffuse electrons or gamma-rays also influences the measurement. To determine the size of the remaining background, off-source measurements are taken. A convenient way to do this is the so-called Wobble observation mode, which permits the simultaneous observation of the signal region and a background region \cite{wobble}. The parameter $\theta^2$, which describes the distance in the telescope camera between the expected source position and the reconstructed source position for each event, defines an ``on'' region and an ``off'' region in the camera. This way, the recorded background events are taken within the same time as the signal events. It is possible to define more than one ``off'' position, to increase the precision of the background measurement. However, to achieve more clarity, only one ``off'' position is used in the application presented here. In this case, the determination of the excess between ``on'' and ``off'' events can be evaluated without further normalization.

\subsubsection{Spectrum reconstruction procedure}
\label{subsubsec:spectrumreconstruction}
During the analysis of MAGIC data with the analysis package MARS \cite{Moralejo:2009p1858}, the recorded shower images are calibrated, cleaned and characterized by so-called image parameters \cite{hillas}. Among these are the \textit{width} and the \textit{length}, describing the root mean square spread of light along and perpendicular to the main axis of the image, the light content of the shower (\textit{size}) and the fraction of light contained in the brightest pixel compared to the total amount of light in the image (\textit{concentration}). Some of these parameters are combined to the \textit{estimated energy}, using the statistical learning method of Random Forest training \cite{Breiman:2001p1823}. This parameter has per construction a very good correlation with the true energy. Similarly, for the best possible gamma/hadron separation, a parameter \textit{hadronness} is built, which describes the probability for each event to be of hadronic origin.

To obtain a differential spectrum with respect to the true gamma-ray energy, an unfolding procedure is applied, using one or several of the observables at hand. The training of the Random Forests and the unfolding procedure require Monte Carlo simulated events, which must have undergone the same analysis procedure up to this point. The simulations used for the analysis of MAGIC data are produced with the air shower simulation package CORSIKA \cite{1998cmcc.book.....H}, followed by the detector simulation \cite{magicmc}. 

At present, the standard MAGIC analysis offers the possibility to perform the reconstruction and unfolding procedure in two subsequent steps. 
First, data events which contain all formerly mentioned parameters are read and cuts are applied on \textit{hadronness}, to select gamma-like events, on the sky coordinates of the recorded events and on $\theta^2$, to separate events from the on- and the off-source measurements. Monte Carlo simulated events are used to determine the acceptance of the detector, the effective area, and the migration matrix for the true energy and the observable \textit{estimated energy}. The product is an energy spectrum of the \textit{estimated energy} and the migration matrix for the chosen binning. 

In a second step, different unfolding algorithms with different regularization methods can be applied, in order to produce a spectrum with respect to the true energy \cite{Collaboration:2007p1826}. Among the offered regularizations, the methods by Bertero \cite{Bertero:1988eh}, Schmelling \cite{schmelling} and Tikhonov \cite{tikhonov1977solutions} can be used. The unfolding is performed using the formerly generated migration matrix. Furthermore, fits of several functions to the obtained spectrum can be performed, taking into account the correlation between the unfolded data points.

 There are several limitations within this procedure. The basis for the unfolding is fixed to the already binned histogram provided in the first step of spectral reconstruction. Thus, an optimization of the binning is not possible during the unfolding process. Furthermore the estimated energy is the only available observable. Additional observable parameters are not accessible anymore, but might yield complementary information. The application of TRUEE offers an alternative approach to the whole reconstruction and unfolding process, which can avoid these unnecessary limitations.

\subsubsection{Application of TRUEE}
As described in section \ref{Sec:truee}, TRUEE offers the usage of up to three observable parameters for the unfolding. It reads the data sample on event-by-event basis instead of ready-made histograms, which leaves the program the freedom to choose an optimal binning for all the observables. The unfolding program also performs the acceptance correction of the detector. Moreover, within the unfolding process, TRUEE can account for the background of the measurement, using a background event sample. In the standard MARS analysis this is done prior to the unfolding.

The fact that individual events are read by the program requires TRUEE to enter the analysis process at an earlier step than the current unfolding tool. This is feasible, as the consideration of the background events does not need to be carried out before the application of TRUEE. Also, the building of preliminary histograms is done inside the program during the unfolding process. Thus, the first step in spectral reconstruction, which has been outlined in \ref{subsubsec:spectrumreconstruction}, is not needed for a TRUEE-based spectrum reconstruction. Still, the applied cuts as well as the determination of signal and background events are required to be carried out prior to the unfolding. Thus, an interface has been implemented to permit TRUEE to enter the analysis workflow of the experiment. The tasks which are conducted by this new interface are outlined in the following.

The program reads the events from data and MC simulation files, applies a cut to exclude hadron-like showers and cuts to choose events which belong to either the signal or the background region of the telescope camera. It creates one output file for signal events, one for background events and a third one for MC events. These files contain all parameters which are relevant for the unfolding, disengaged from the MAGIC data file tree structure, as even-level branches. Furthermore, the program reads basic information about the produced MC events and stores them into an extra tree in the MC file. From the data sample, the effective observation time is extracted and added to the signal output file in an additional tree. This additional information are needed for the acceptance correction which is done within TRUEE as well.

After the unfolding process, a script file is used to extract the solution with the best combination of parameters and to apply a fitting algorithm which takes into account the correlation between the bins of the unfolded spectrum. Quoting such a fit result, in addition to the unfolded data points, is common for the presentation of energy spectra in astroparticle physics, as it facilitates the comparison of results from different analyses. The corresponding fit function can be selected by the user among several choices (power law, broken power law, etc.).

\subsubsection{Choice of observables}
\begin{figure*}
\begin{minipage}[b]{0.5\linewidth}
\begin{center}
\includegraphics[width=3.in]{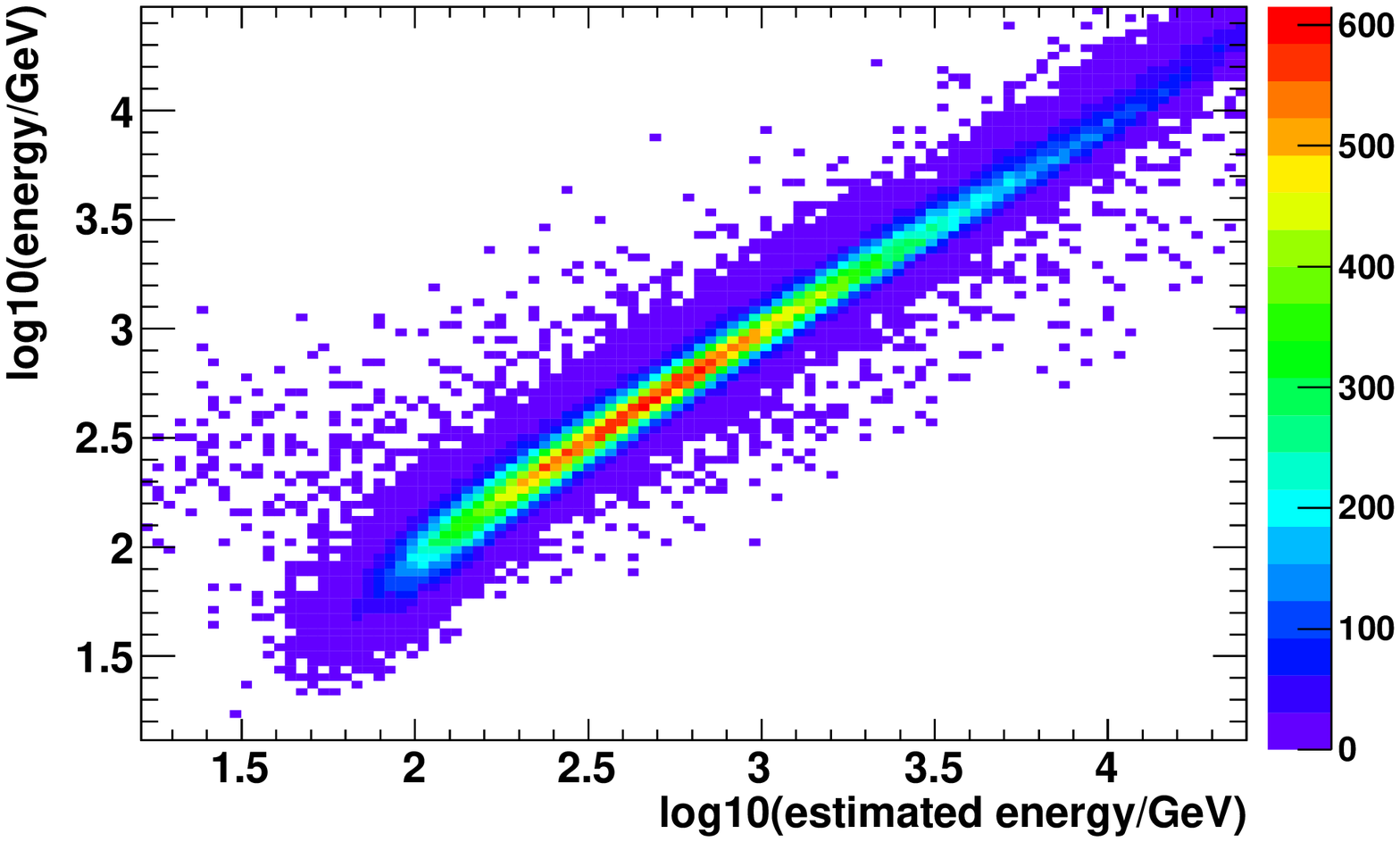}
\vspace*{-1mm}
\includegraphics[width=3.in]{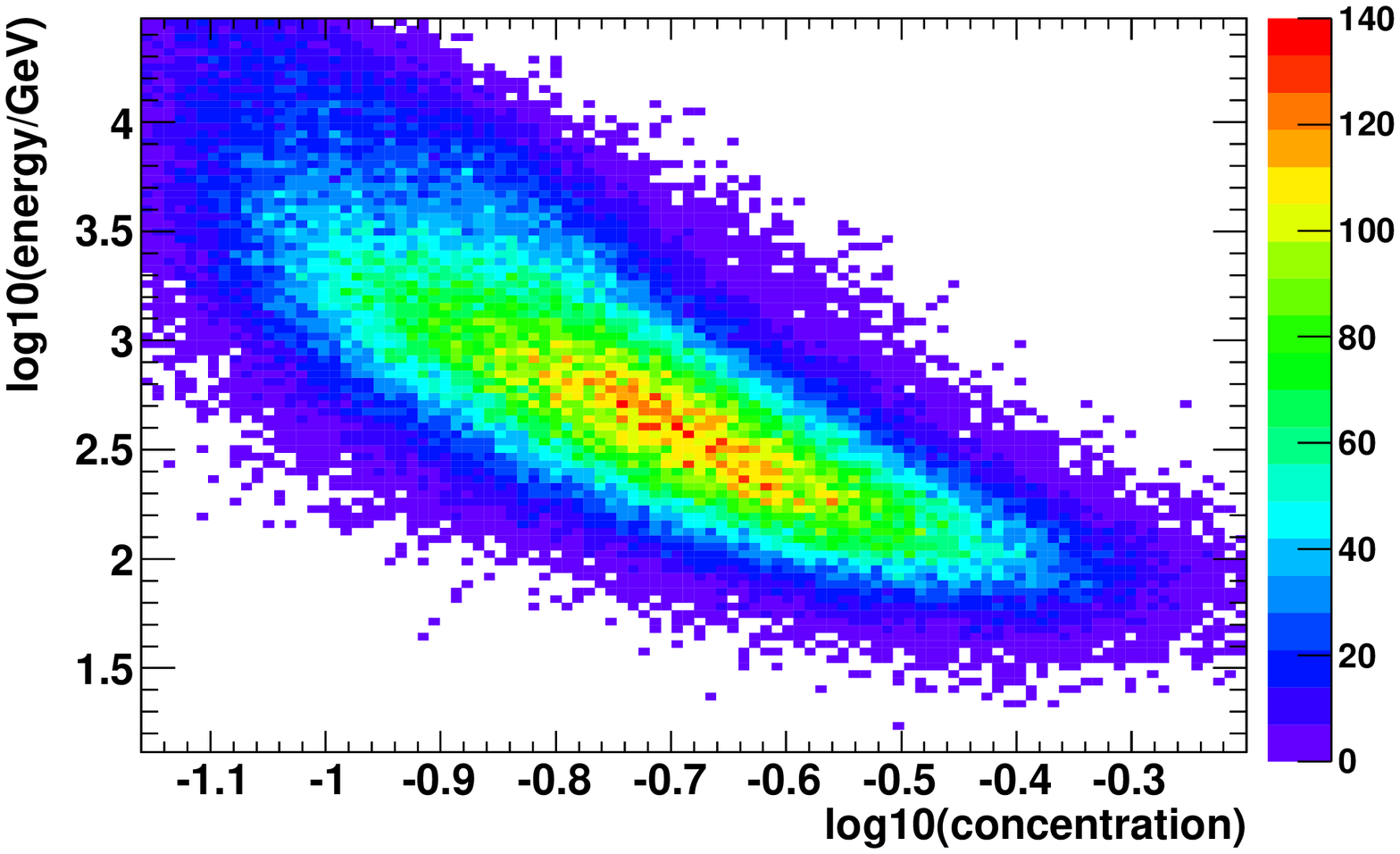}
\vspace*{-1mm}
\includegraphics[width=3.in]{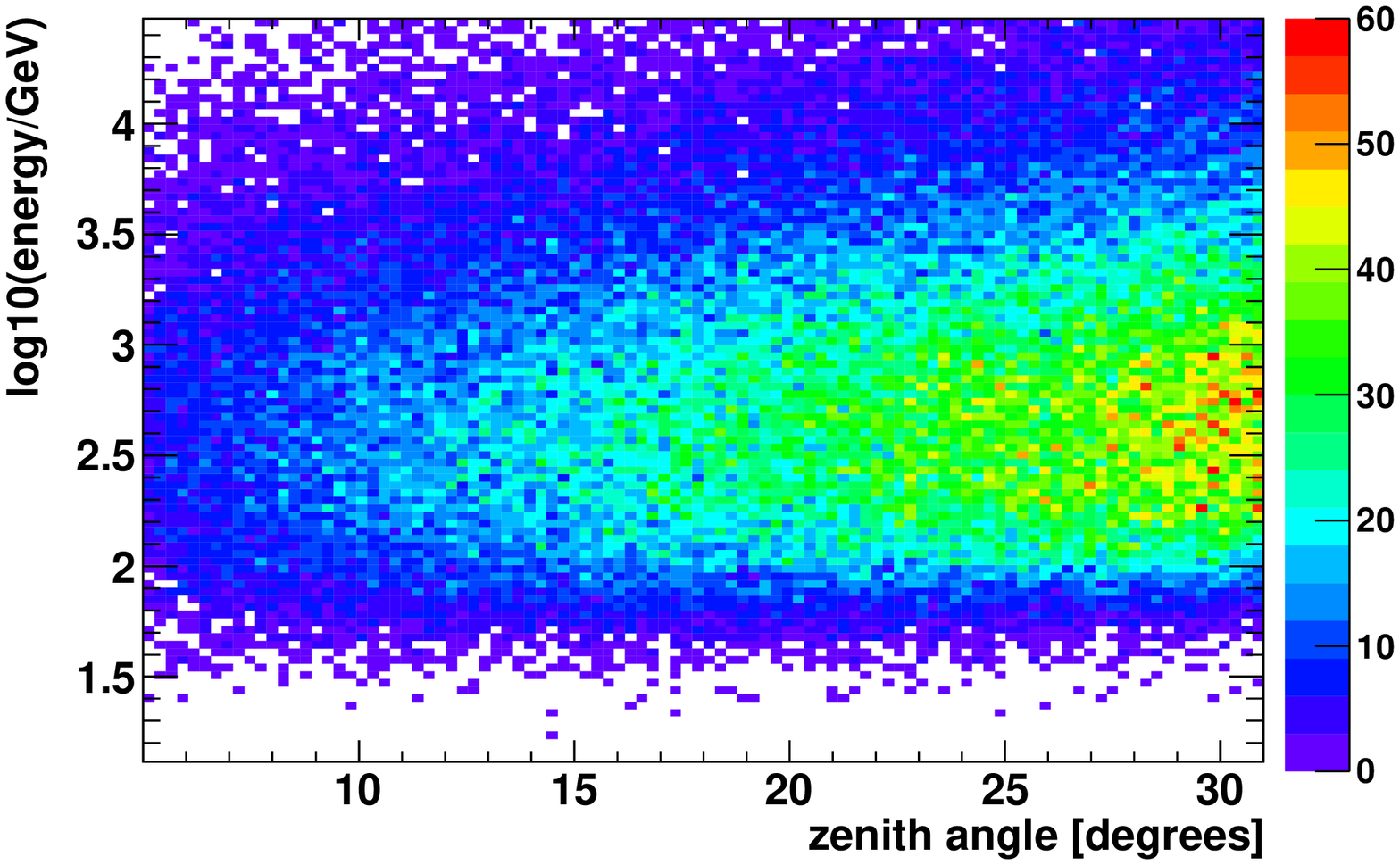}
\vspace*{-1mm}
\end{center}
\end{minipage}
\hspace*{1mm} 
\begin{minipage}[b]{0.5\linewidth}
\begin{center}
\includegraphics[width=3.in]{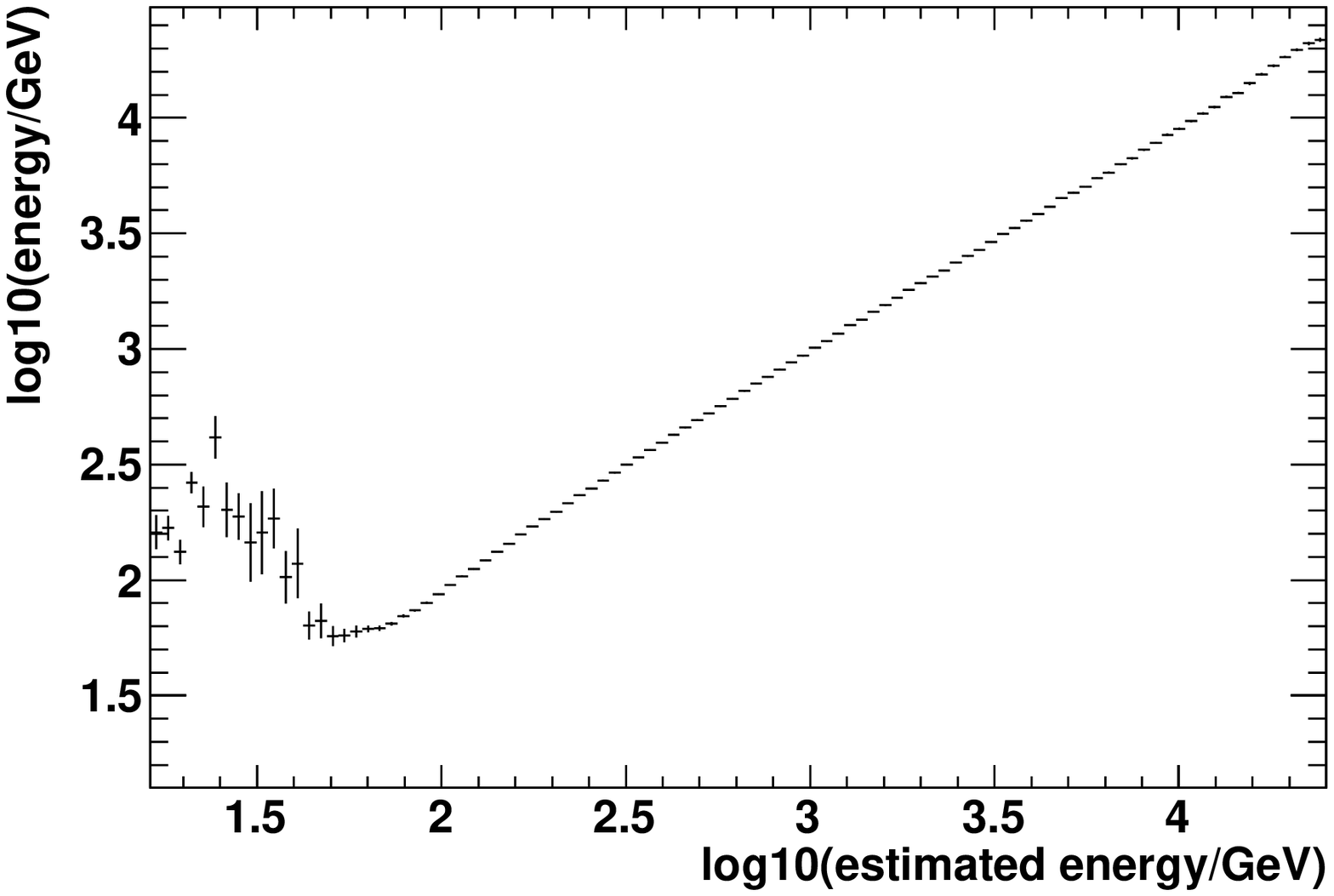}
\vspace*{-1mm}
\includegraphics[width=3.in]{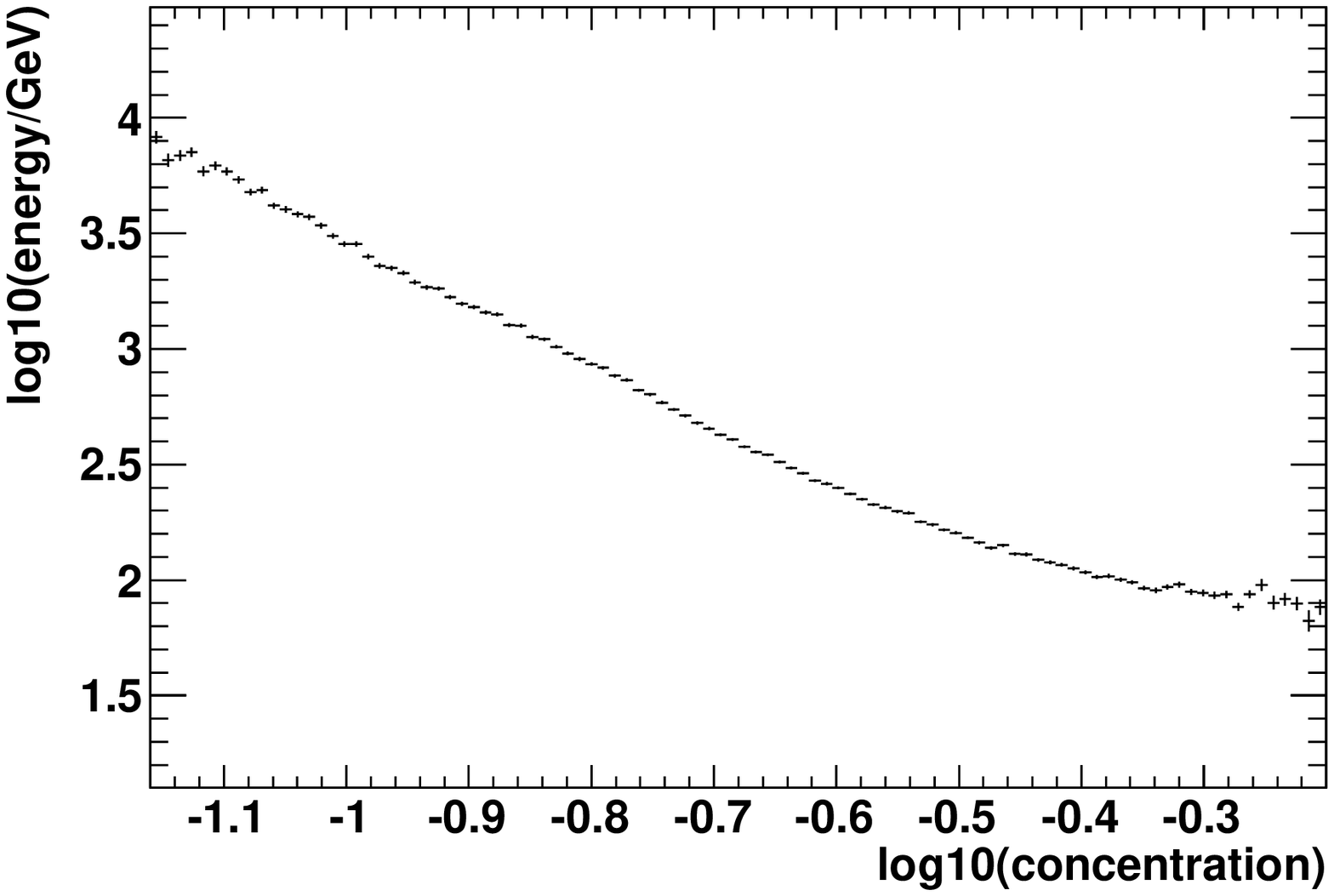}
\vspace*{-1mm}
\includegraphics[width=3.in]{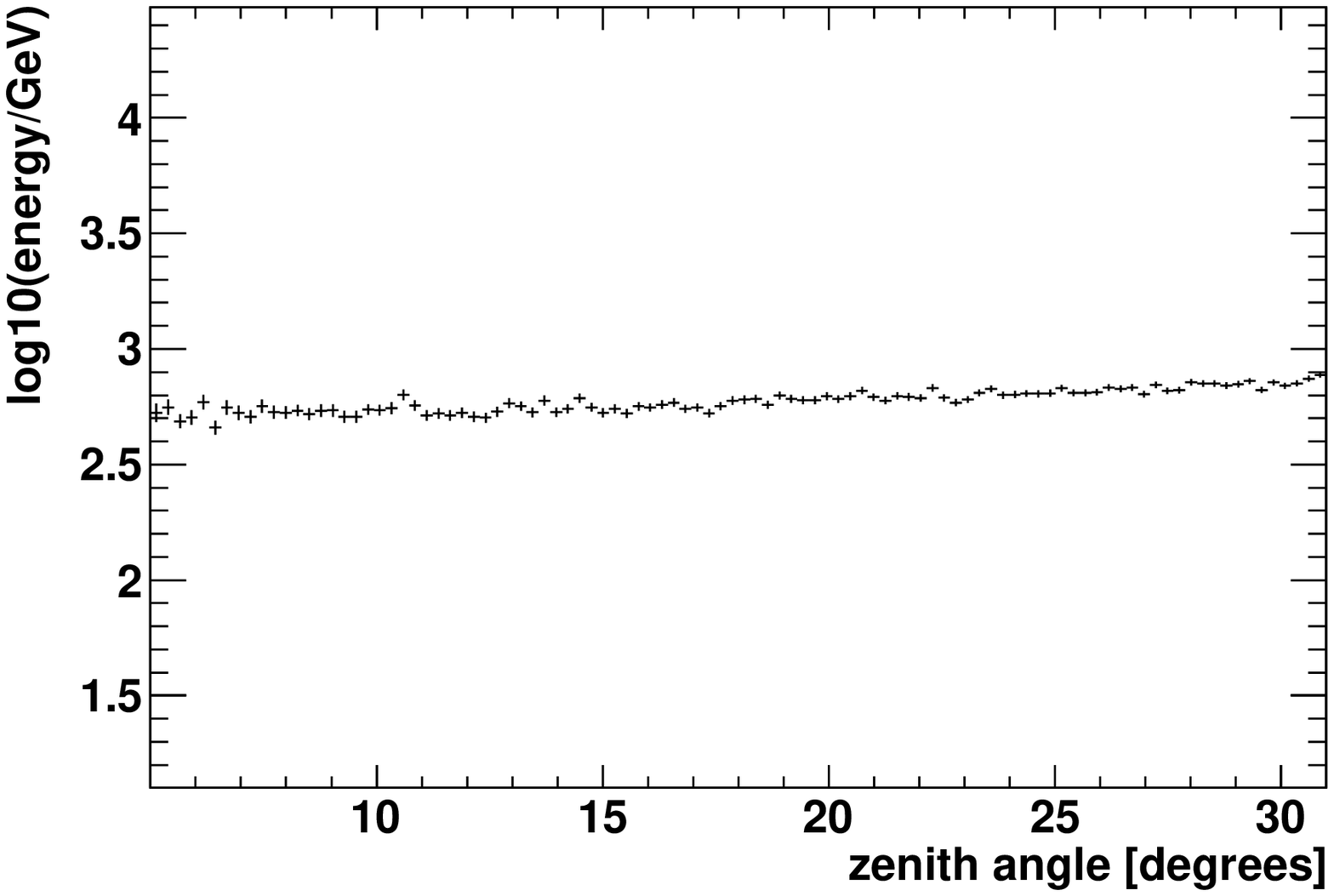}
\vspace*{-1mm}
\end{center}
\end{minipage}
\caption{\label{fig:magiccorellation}Correlation between the energy and the observational parameters used for the application of TRUEE in the MAGIC analysis. Shown are scatter plots of events (left) and the related profile histograms (right). An optimal correlation is present in a monotonically changing profile function with small uncertainties. The density is displayed in color code.}
\end{figure*}
For the unfolding of MAGIC data the space of observable parameters has been investigated. The set of parameters which has proven to deliver good results are:
\begin{itemize}
\item The \textit{estimated energy}, a combination of observables which is gained by random forest regression and correlates very well with the true energy, has been the scaffolding of the current MAGIC unfolding and is a fruitful contribution also for the unfolding with TRUEE.
\item The parameter \textit{concentration}, which describes the light content ratio of the brightest pixel compared to the surrounding ones, shows a clear correlation with the true energy.
\item As a third parameter the \textit{zenith angle} is an important input for an unfolding with TRUEE. Even though it does not show a good correlation with the energy, it influences the image of each event in the camera, so that events with the same energy look different if they have been taken at different zenith angles.
\end{itemize}
The correlation of each observable parameter with the true energy is shown in Fig.~\ref{fig:magiccorellation}.

\subsubsection{Acceptance correction}
\label{subsubsec:acceptance}
The data events, which are read by TRUEE, are only those events which triggered the telescopes and which survived the analysis cuts. Similarly, the MC set only comprises events which remain after the selection by a simulated trigger and the same analysis cuts. In other words, the measurement is affected by a loss of events compared to the initially arriving particles. The ratio between the distributions of the surviving and the arriving particles is given by the acceptance of the measurement process. As explained in \ref{Subsec:acceptance}, TRUEE can deduce this acceptance and apply an appropriate correction during the unfolding, in order to get the true distribution in the sought quantity.
For MAGIC data, this generally is the differential flux of gamma-ray particles, i.e.~the number of particles per unit area, time and energy. Thus, the distribution of the MC has to be expressed in the same way. The area in which the MC events are generated is given by a circle whose radius is the so-called \textit{maximum impact parameter} $r$. As MC events do not have a density in time, this factor has to be determined by the following consideration. To obtain the actual factor between the number of data events collected within the actual observation time and the initial flux from the simulations, MC and data events have to be related to each other. For this reason the MC distribution is normalized to the effective observation time $T_{obs}$ which the real data sample was collected in.

Adding this information, the normalization constant of the MC distribution can be obtained from the number of generated particles $N_{gen}$, the energy range which has been simulated ($E_{min}$ to $E_{max}$) and the spectral index $\gamma$, using
\begin{eqnarray}
	\label{eq:acceptancecorr}
	\frac{d^3N}{dE \:dA\: dt}= C \cdot \left(\frac{E}{1\,GeV}\right)^{-\gamma}.\label{eq:MCdistr}
\end{eqnarray}
Integrating yields the number of generated events,
\begin{eqnarray}
	N_{gen}=\int\limits_{E_{min}}^{E_{max}} \int\limits_{A} \int\limits_{T_{obs}} \frac{d^3N}{dE\:dA\:dt}dE\:dA\:dt,
\end{eqnarray}
which implies that
\begin{eqnarray}
	C &=& \frac{N_{gen}}{{r}^2 \cdot \pi \cdot T_{obs}} \nonumber \\
 &&\cdot  \frac{(-\gamma +1) }{\left(\frac{E_{max}}{1\,GeV}\right)^{-\gamma +1} -\left(\frac{ E_{min}}{1\,GeV}\right)^{-\gamma +1}}.
\end{eqnarray}
In the case of unfolding a MC sample, the flux of particles per time is not a meaningful parameter, as neither the auxiliary MC sample nor the target sample have such a time density information. So in that case, a distribution of particles per area and energy is used.

\subsubsection{Unfolding of MC spectra}
The unfolding of MC data is presented in several steps. First, a test unfolding is performed. In a second step, a different MC sample is used as pseudo data. These unfolding procedures handle only the distribution of events which remain after the trigger simulation and cuts set during the analysis. Subsequently, two examples of an unfolding with an applied correction for the acceptance of the detector is shown, which results in a differential flux spectrum, i.e.~in the case of MC the initial number of particles per energy and area which have been generated in the MC simulation. The used MC samples are summarized in Tab.~\ref{tab:magicmcsamples}. The features specified there are the spectral index of the generated power law distribution, the range in zenith angle which is covered by the simulated events and the \textit{maximum impact parameter} $r$, which specifies the area over which the generated events are distributed. Furthermore the number of primarily generated events is given as well as the number of events which survive after triggers and analysis cuts.

\begin{table} [!thh]
	\begin{tabular}{|c|c|c|c|}
		\hline
		MC Sample & A  & B & C\\
		\hline
		Spectral index $\gamma$& 1.6  &2.6 & 2.6\\
		Zenith angle range & 5$^\circ$ - 35$^\circ$ & 5$^\circ$ - 35$^\circ$ & 5$^\circ$ - 35$^\circ$\\
		Impact parameter & 350 m& 350 m & 350 m\\
		No.~generated & 7\,893\,000 &10\,000\,000 & 40\,000\,000\\
		No.~residual & 272\,283 &24\,444 & 100\,224\\
		\hline
\end{tabular}

\caption{\label{tab:magicmcsamples}Summary of the Monte Carlo samples which are used during the unfolding of MAGIC Monte Carlo and data events.}
\end{table}

\begin{figure}[!thh]
		 \centerline{\includegraphics[width=3.in]{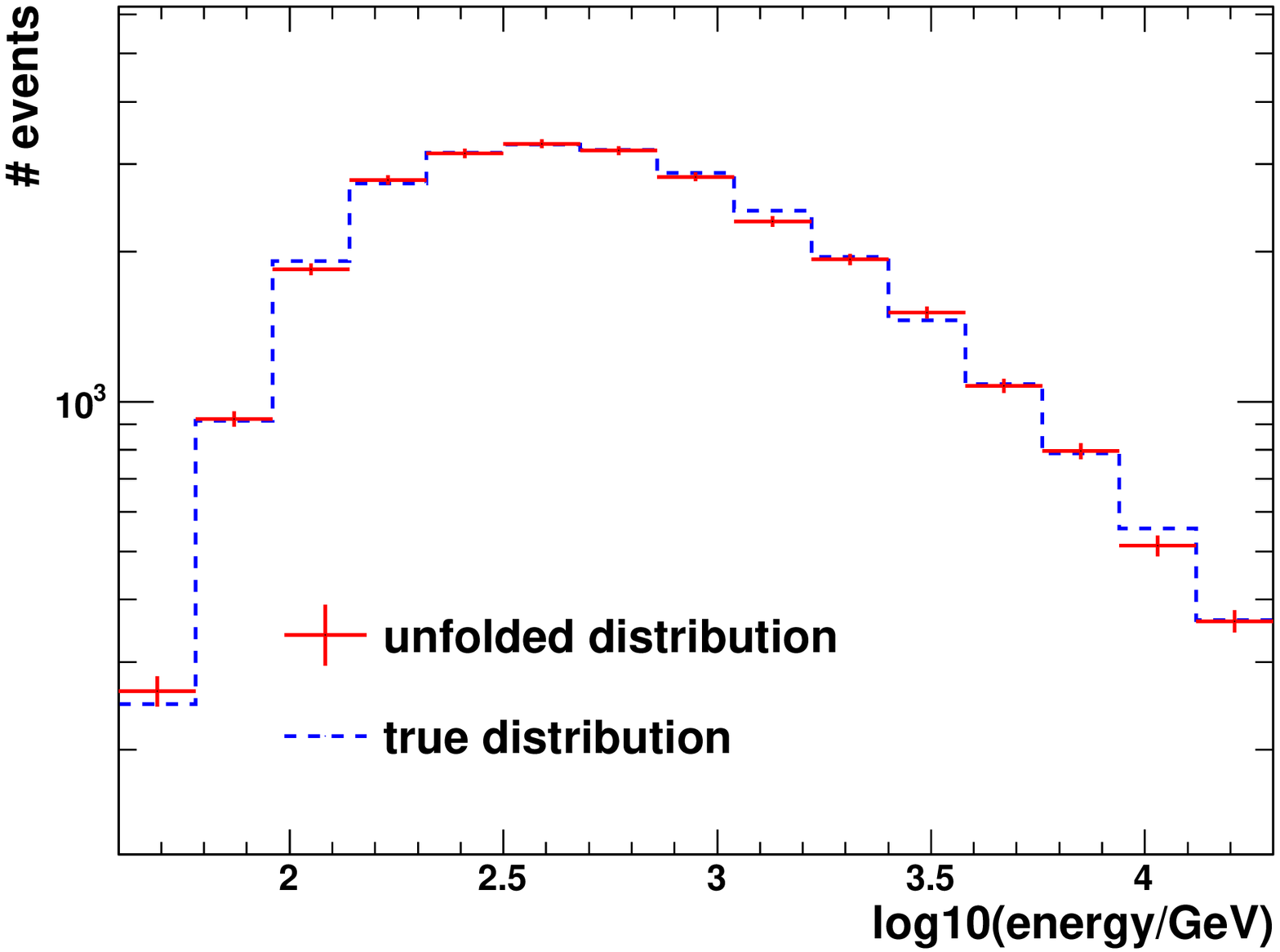}}
		 \centerline{\includegraphics[width=3.in]{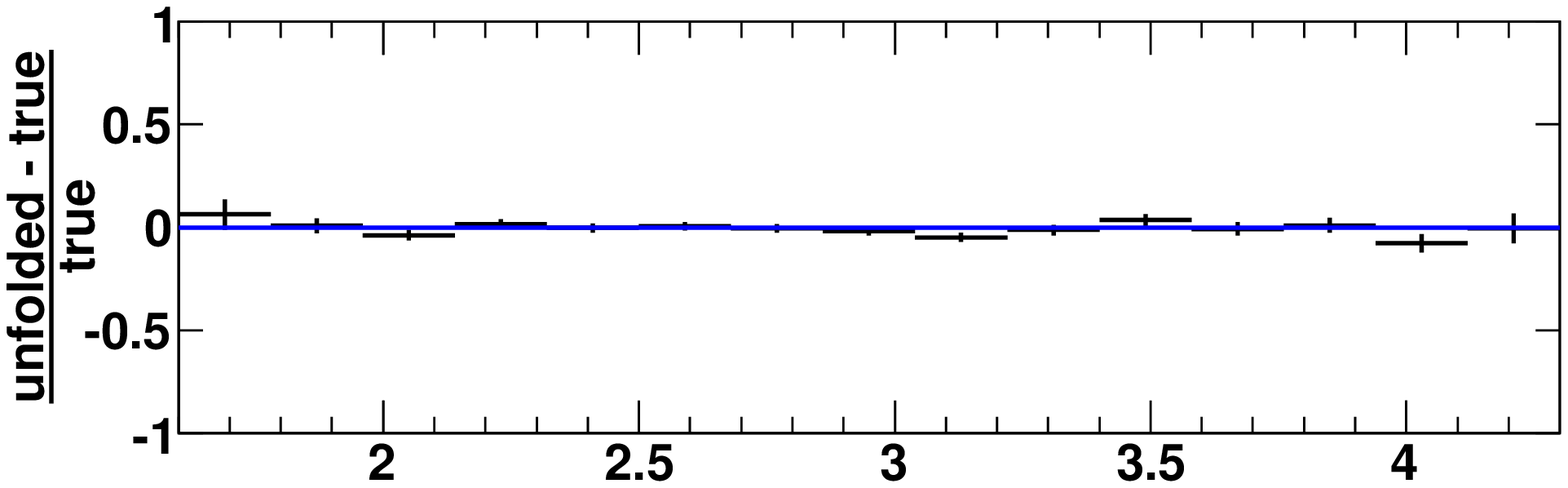}}
	\caption{Event distribution obtained with an unfolding of MC events in TRUEE's test mode. The unfolding result (red points with error bars) is compared to the true distribution known from the MC (blue/dashed curve).}
	\label{fig:testunfoldingcomp}
	\end{figure}

Following the above scheme, first of all a test unfolding is performed, using only MC sample~A. $10\,\%$ of the MC is taken as pseudo data to be unfolded, and $90\,\%$ are used for the response matrix and acceptance calculation. 
A wide range of unfolding parameters is probed and the combination which delivers the smallest inter-bin-correlation is chosen. For the analysis presented here, which results in a spectrum with 16 bins, these are 21 knots and 13 degrees of freedom. The resulting MC spectrum can be seen in Fig.~\ref{fig:testunfoldingcomp}, where additionally the true distribution of events is shown. This direct comparison of the unfolded and the true distribution permits to verify the goodness of the choice of the used unfolding parameters. 

A successful test unfolding delivers a good reproduction of the input spectrum and information about which parameter combinations give reliable results. However, in this case the MC and the pseudo data events follow the same distribution as they stem from the same MC sample. As the simulated and the real distribution of the data are in general not equal, the performance for different distributions in MC and pseudo data needs to be investigated. For this purpose a second MC unfolding is carried out, this time applying MC sample B as pseudo data, while the whole sample A is used to calculate the response matrix. The two MC sets show different distributions in energy, such that the spectral indices differ by 1.0. The number of generated events in sample B is higher, but due to the steeper spectrum and the decrease of the trigger efficiency towards low energies, the final sample is one order of magnitude smaller than MC sample A. The ratio of events between data and MC events of $\sim10$ is also desirable for the unfolding of real data.

\begin{figure}[!thh]
	\centerline{\includegraphics[width=3.in]{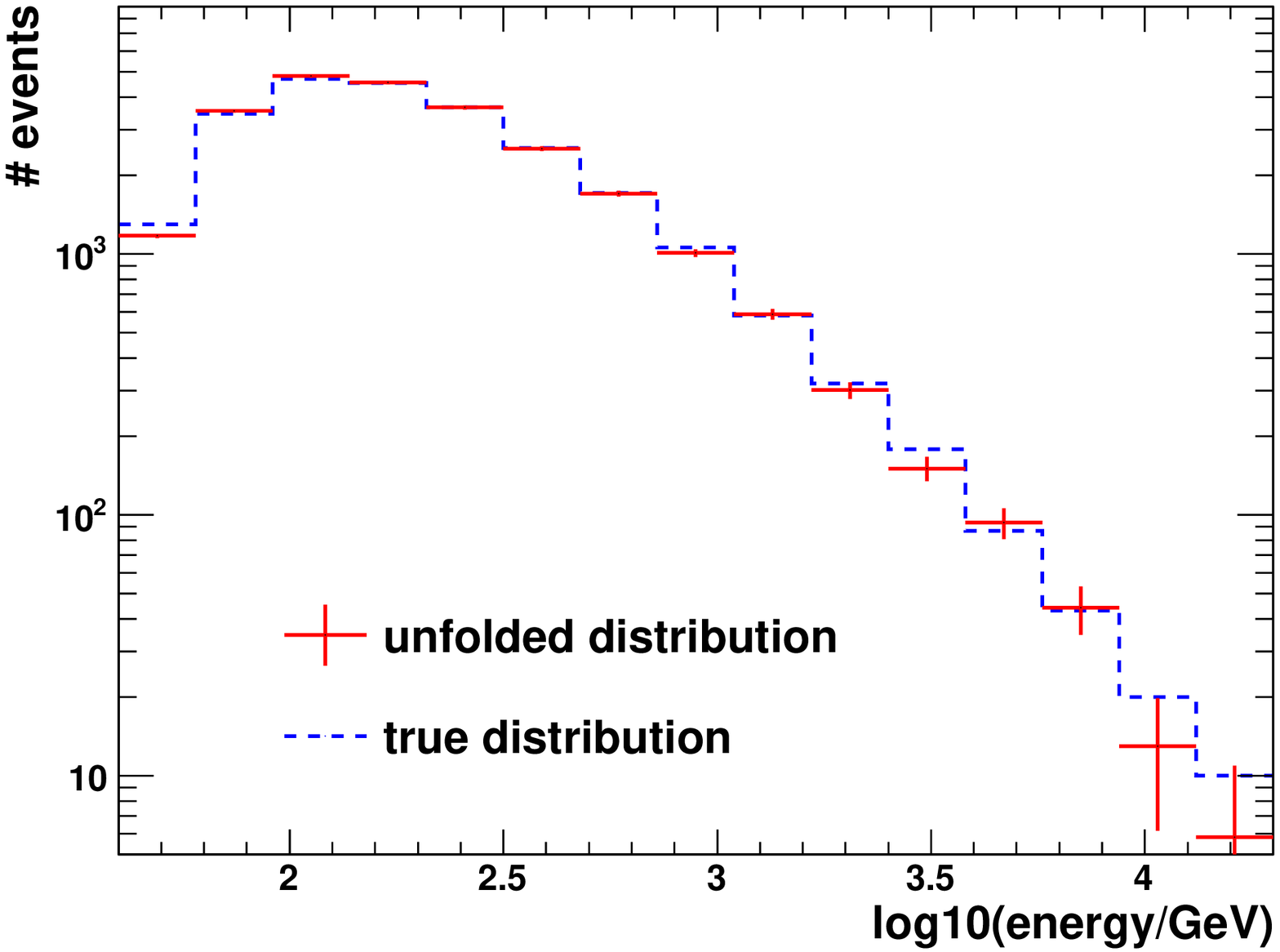}}
\centerline{\includegraphics[width=3.in]{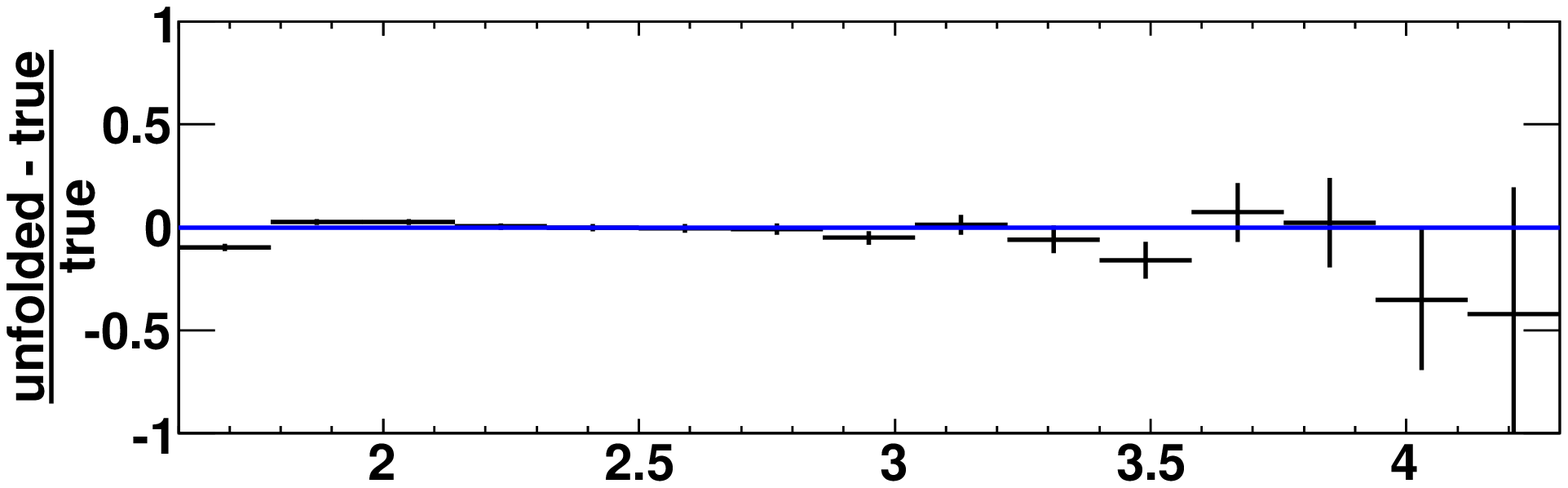}}
	\caption{Unfolded event spectrum of MC events. MC sample B with spectral index $\gamma_B=2.6$ has been unfolded as pseudo data, applying MC sample A with spectral index $\gamma_{A}=1.6$. The unfolded points of the event spectrum (red points with error bars) and the original distribution (blue/dashed curve) are shown.}
	\label{fig:magicmccomparison}
\end{figure}

The unfolding process is carried out with the same binning of observable parameters and combinations of unfolding parameters as the test unfolding shown above. The result can be seen in Fig.~\ref{fig:magicmccomparison}.

After the unfolding of triggered event distributions, the unfolding of a MC sample with applied acceptance correction will be shown. For this purpose, MC sample B serves as pseudo data again, while sample A forms the MC sample for the determination of the response matrix. Additionally, the initial distribution of sample A is given as 
\begin{eqnarray}
\label{eq:MCinitialdistA}
\frac{d^2N}{dEdA}&=& \frac{N_{gen}}{{r}^2 \cdot \pi} \cdot (-\gamma +1) \nonumber \\
&& \cdot \frac{\frac{E}{1\, GeV}^{-\gamma}}{\left(\frac{E_{max}}{1\,GeV}\right)^{-\gamma +1} - \left(\frac{E_{min}}{1\,GeV}\right)^{-\gamma +1}} , 
\end{eqnarray}
with $E_{min}=10$\,GeV and $E_{max}=30\,000$\,GeV. For the remaining quantities see Tab.~\ref{tab:magicmcsamples}.

\begin{figure}[!thh]
            \centerline{\includegraphics[width=3.in]{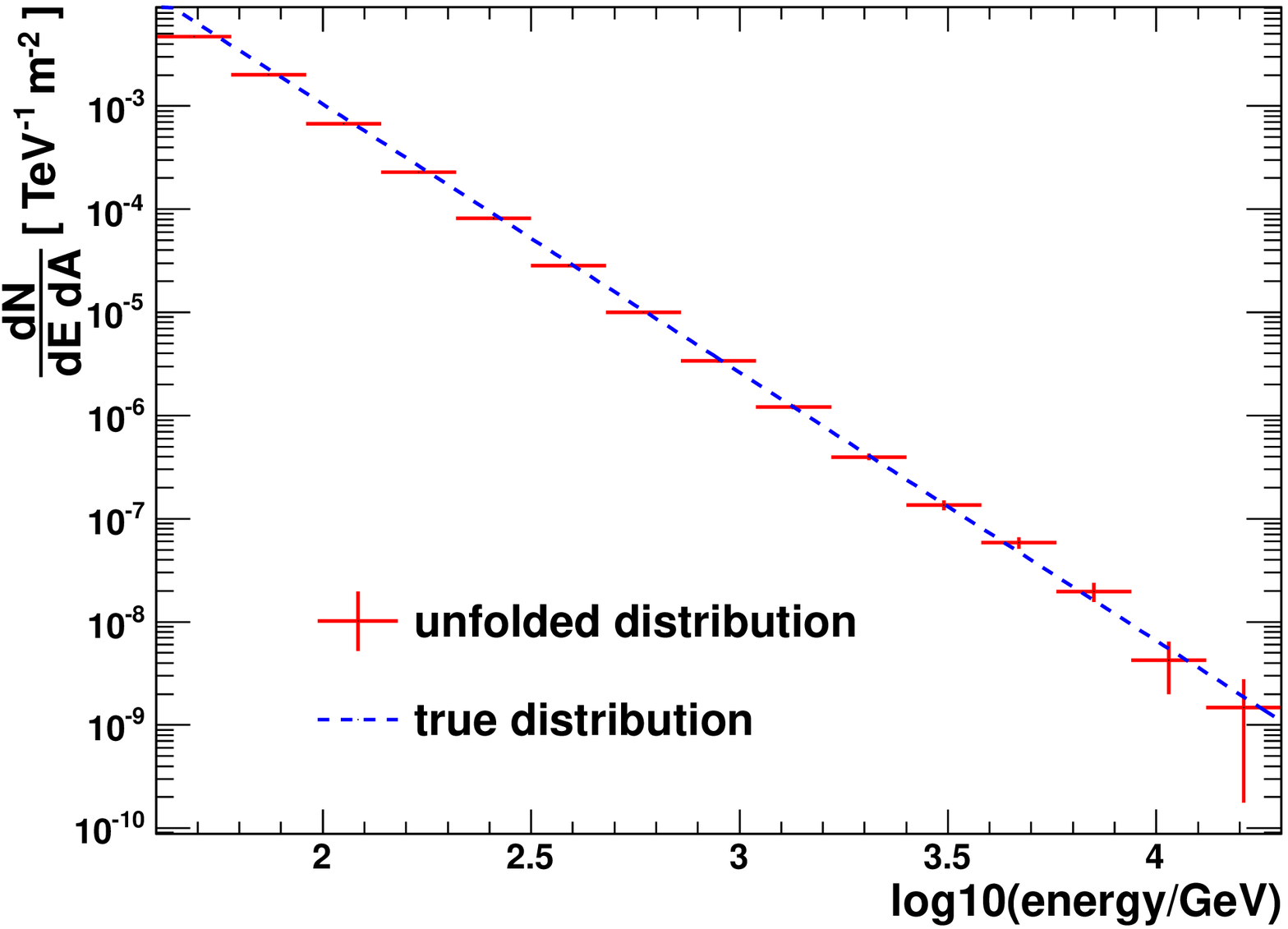}}
	    \centerline{\includegraphics[width=3.in]{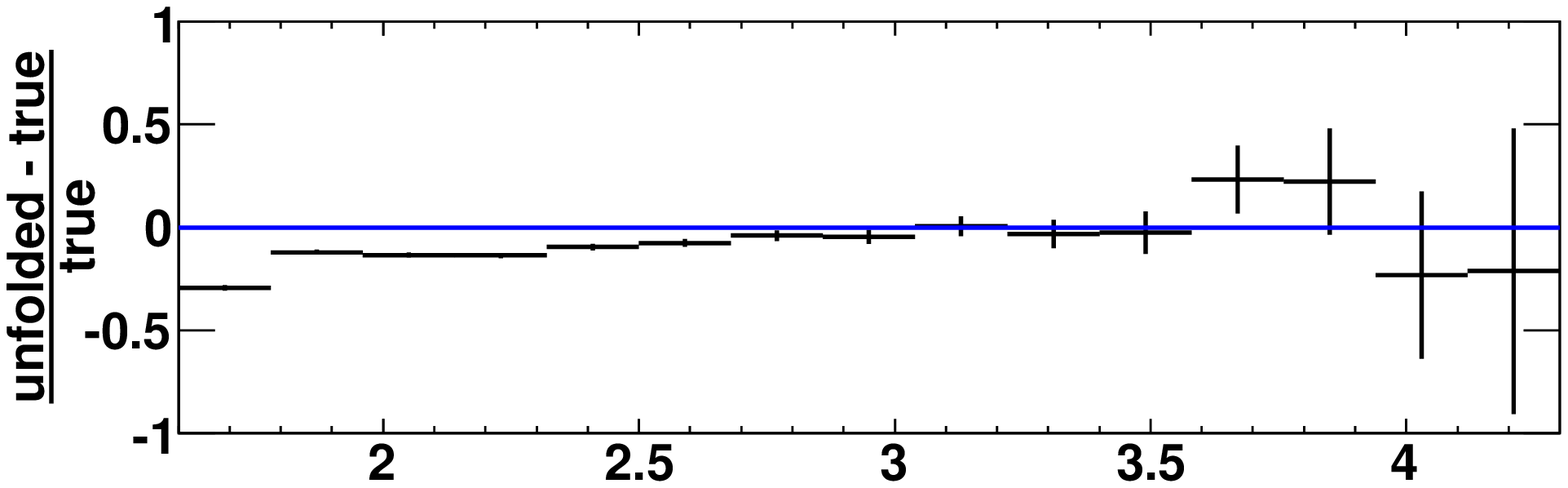}}

	\caption{Unfolding of MC simulations as pseudo data with built-in acceptance correction of TRUEE. MC sample~B is unfolded as pseudo data, while MC sample~A serves as MC in the unfolding. The red points with error bars represent the result of the unfolding. The blue solid line shows the true initial MC distribution.}
	\label{fig:magicmcacceptance}
\end{figure}

The unfolding itself is performed with the same binning of the observables and with the same unfolding parameters. Figure~\ref{fig:magicmcacceptance} shows the unfolded distribution and the initial MC function. It has to be noted that, while a good agreement is achieved at intermediate energies, the distribution at lower energies appears to be systematically underestimated. This effect is caused by the fact that the acceptance correction refers to the center of gravity within each bin of the initial MC distribution. For large differences between the spectrum of the MC sample used to determine the response matrix and of the spectrum obtained by the unfolding procedure, the relative shift between the centers of gravity of the two distributions is not negligible anymore. \\
However, this effect can be corrected, if - in the case of such discrepancies in the spectra - a second step of unfolding with acceptance correction is applied, using a re-weighted MC spectrum which is more similar to the result of the first step. 

\begin{figure}[!thh]
	    \centerline{\includegraphics[width=3.in]{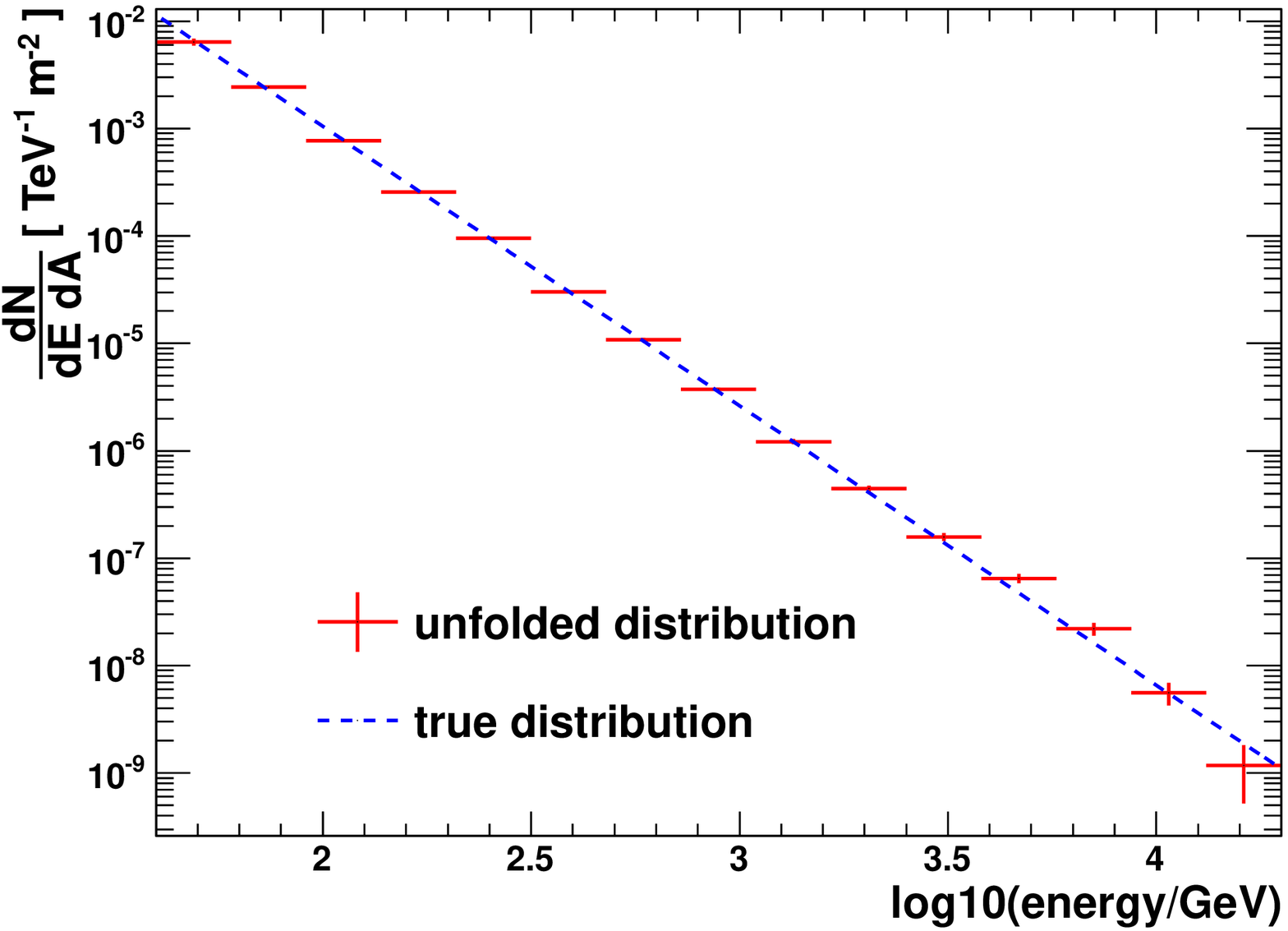}}
	    \centerline{\includegraphics[width=3.in]{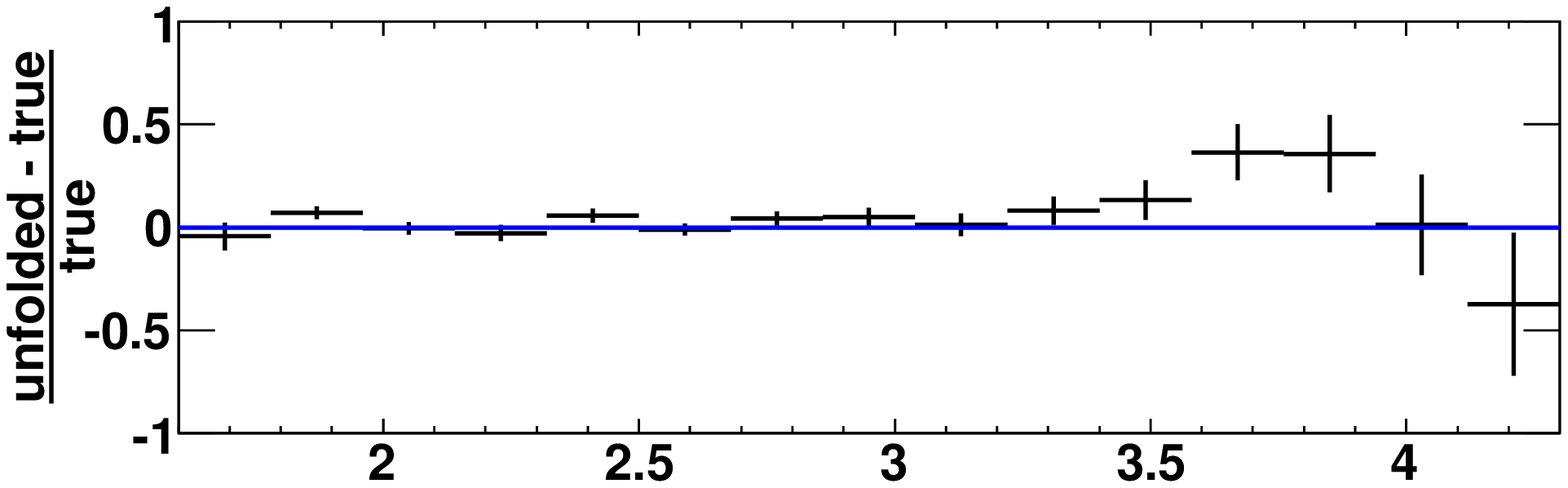}}

	\caption{Unfolding of MC simulations as pseudo data with built-in acceptance correction of TRUEE. Unlike in Fig.~\ref{fig:magicmcacceptance}, the pseudo data sample MC~B is unfolded using MC sample~C, which features the same spectral slope. The result of the unfolding is given by red points with error bars. The blue solid line represents the true initial MC distribution.}
	\label{fig:magicmcacceptancenew}
\end{figure}

For the study presented here, the unfolding of the pseudo data sample with acceptance correction is repeated using MC sample C (see Tab.~\ref{tab:magicmcsamples}), which shows the same spectral index as the pseudo data. The result can be seen in Fig.~\ref{fig:magicmcacceptancenew}. Obviously, the formerly seen discrepancies at low energies do not appear in this case.

Still, the unfolding itself is only slightly affected by this dependency. It delivers good results for the event spectra, also for this significant difference in the spectral indices, as can be seen in the  Fig.~\ref{fig:magicmccomparison}. The remaining discrepancy at very low energies disappears after the above mentioned correction.


\subsubsection{Unfolding of a MAGIC data sample}
After the successful application of TRUEE in the MAGIC unfolding of MC spectra, a proof of principle on real telescope data is given in the following. For this purpose the standard candle of gamma-ray astronomy, the Crab Nebula, serves as an exemplary source. A test data sample which is described below is analyzed with both the standard MAGIC analysis chain and the new chain including TRUEE. Finally a comparison of the results is shown. We would like to state at this point that the presented analysis is not optimized for extracting any results regarding the physics of the observed source or the telescope performance. It only serves as an example of the compatibility of the two analyses. Studies on the performance of the MAGIC stereo system can be found in \cite{performancepaper}.

The data sample comprises 7.3 hours of Crab Nebula observations taken with the MAGIC telescopes. The data have been taken in Wobble observation mode.

The preparation of the data, including the conversion of the extracted charge into the number of photons at the photodetector, the cleaning of the shower images and the determination of image parameters to the light distributions are identical for both analyses. The standard MAGIC analysis and the TRUEE-based analysis diverge at the point where both the data and the MC are preprocessed such that the events are all characterized by image parameters and are assigned an \textit{estimated energy} and a \textit{hadronness}. 

In the current MAGIC analysis, a standard analysis following \cite{performancepaper} has been used to derive cuts for the separation of gammas and hadrons and for the determination of the ``on'' and the ``off'' event sample, using the standard spectrum reconstruction tool. The obtained cuts are applied on the MC and on the data sample. MC sample A is used to determine the effective area of the measurement and to build the migration matrix. The resulting spectrum of differential flux vs. estimated energy is unfolded using the current MAGIC tool and the generated migration matrix. Several executions with different regularization methods have been performed, leading to compatible results.

Following the analysis chain presented in this paper, the event files are processed by the aforementioned interface. The applied analysis cuts are the same as the ones used in the standard MAGIC analysis example. TRUEE is performed using also MC sample A to calculate the migration matrix and to obtain the overall acceptance.  Events which are sorted into the off-source measurement sample are given to TRUEE as background events. For the case of one ``off'' region, no normalization in terms of weighting has to be applied to these events as in Wobble mode, the on- and off-source measurements are taken simultaneously. The chosen set of unfolding parameters, namely number of bins, number of knots and number of degrees of freedom, is the one which has proven to deliver good results during the MC based unfolding procedures discussed before. 

\begin{figure}[!thh]
\centerline{\includegraphics[width=3.in]{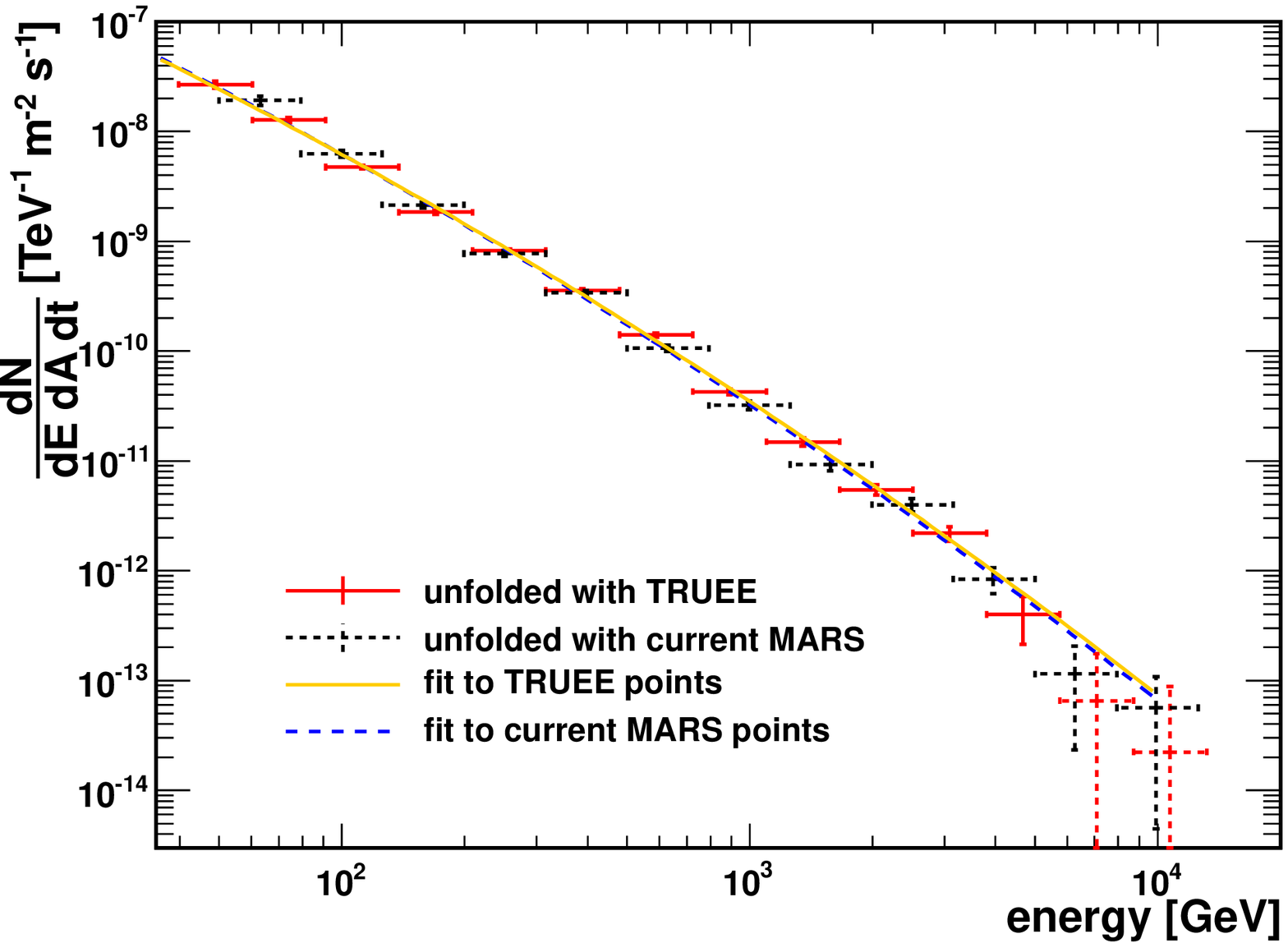}}
\centerline{\includegraphics[width=3.in]{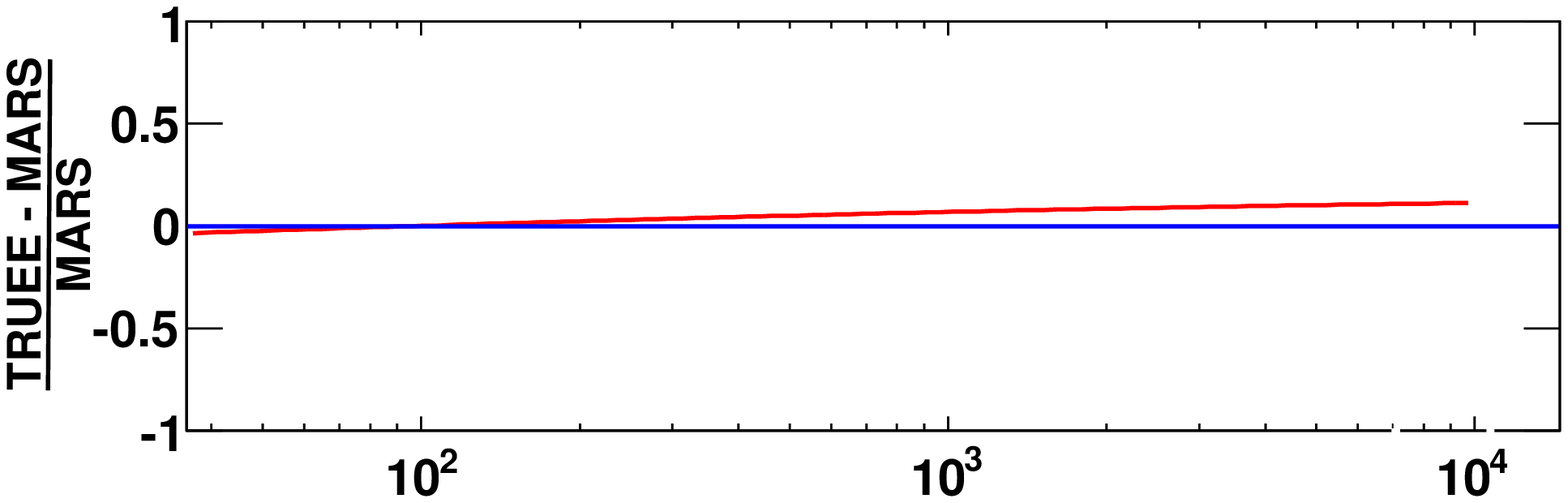}}
	
	\caption{The upper panel presents the unfolded energy spectrum of the Crab Nebula, produced by the example analyses presented here. Shown are results obtained with the current MAGIC unfolding (black/dashed points with error bars) and TRUEE (red/solid points with error bars). Also shown are fits of curved power laws to the unfolded data points. The fit to the standard MARS unfolding is shown in blue/dashed, the fit to TRUEE-unfolded points is depicted in orange/solid. In the bottom panel, the relative deviation of the two fitted functions with respect to the energy is shown.}
	\label{fig:magicdatacomparison}
\end{figure}

The comparison of the results, produced with the two analyses, are shown in Fig.~\ref{fig:magicdatacomparison}. The two spectra show a good agreement, with deviations below $11\,\%$.

\subsubsection{Verification}
\label{sec:verificationMagic}
\begin{figure*}
\begin{minipage}[b]{0.5\linewidth}
\begin{center}
\includegraphics[width=3.in]{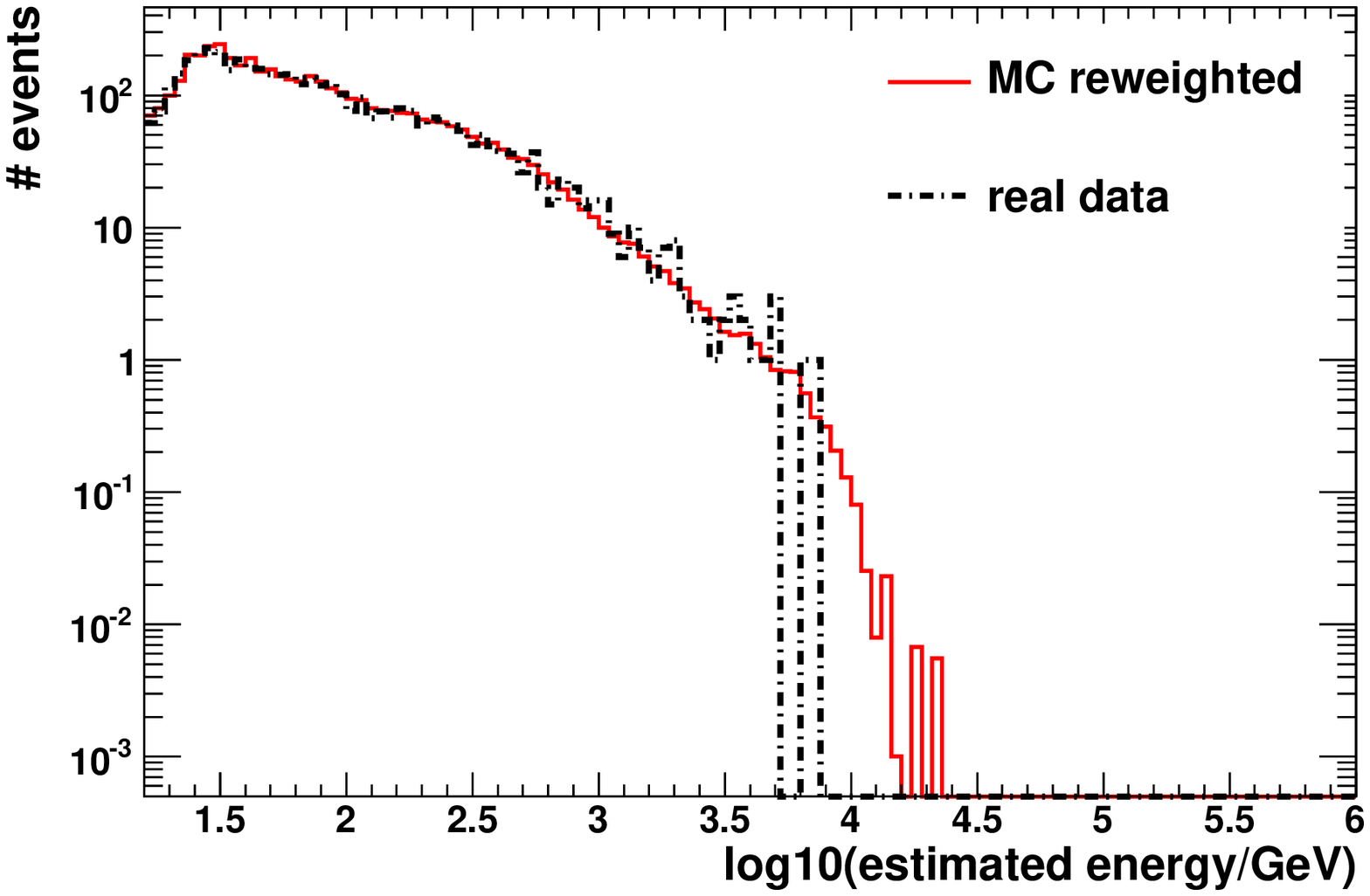}
\vspace*{-1mm}
\end{center}
\end{minipage}
\hspace*{1mm}
\begin{minipage}[b]{0.5\linewidth}
\begin{center}
\includegraphics[width=3.in]{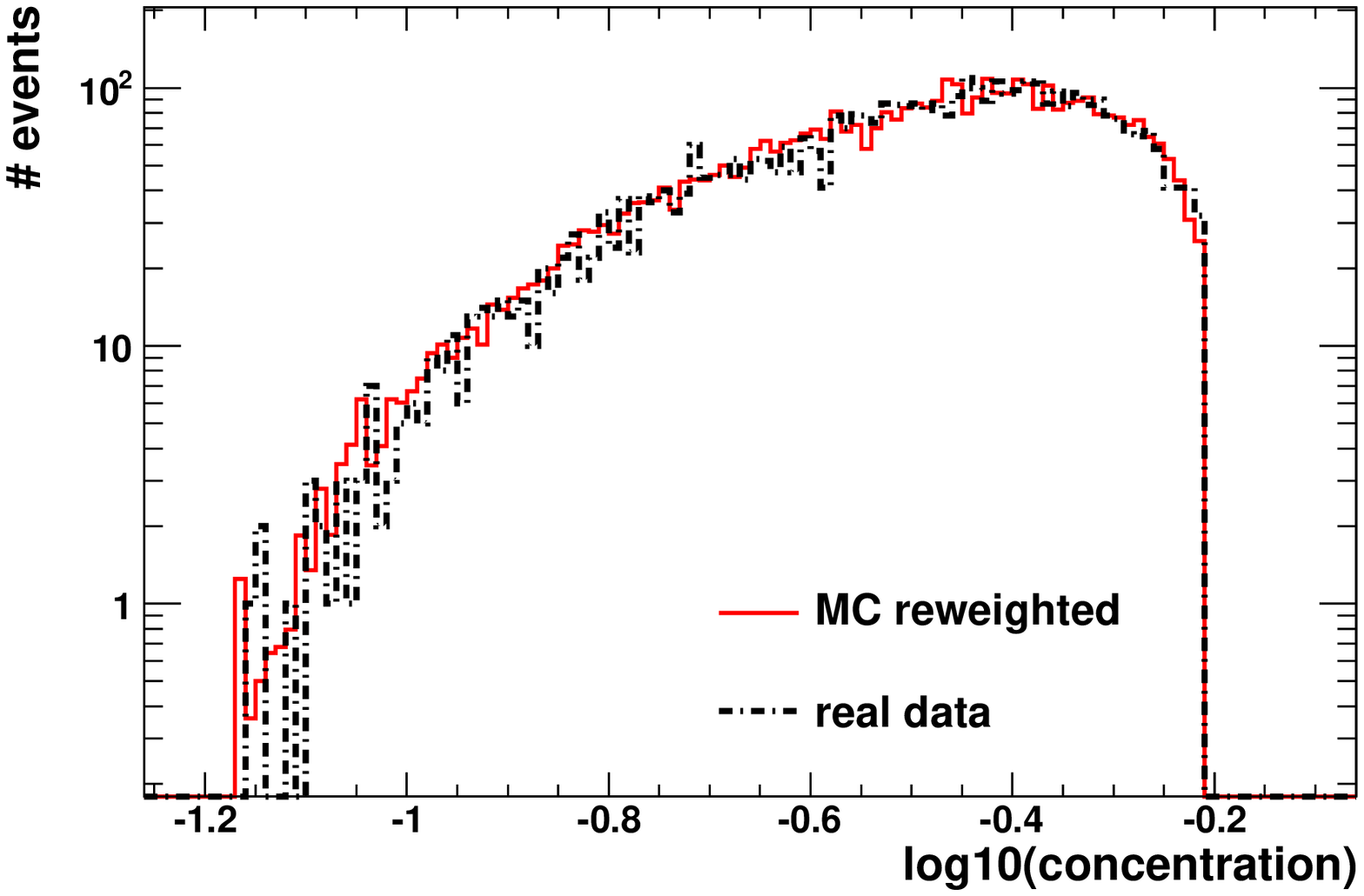}
\vspace*{-1mm}
\end{center}
\end{minipage}
\caption{\label{fig:magiccheckunfoldingvariables}Comparison of distributions in observable parameters which have been used during the unfolding fit. Shown are the estimated energy and concentration distributions for real data (black/dot-dashed) and the re-weighted MC (red/solid).}
\end{figure*}

\begin{figure*}
\begin{minipage}[b]{0.5\linewidth}
\begin{center}
\includegraphics[width=3.in]{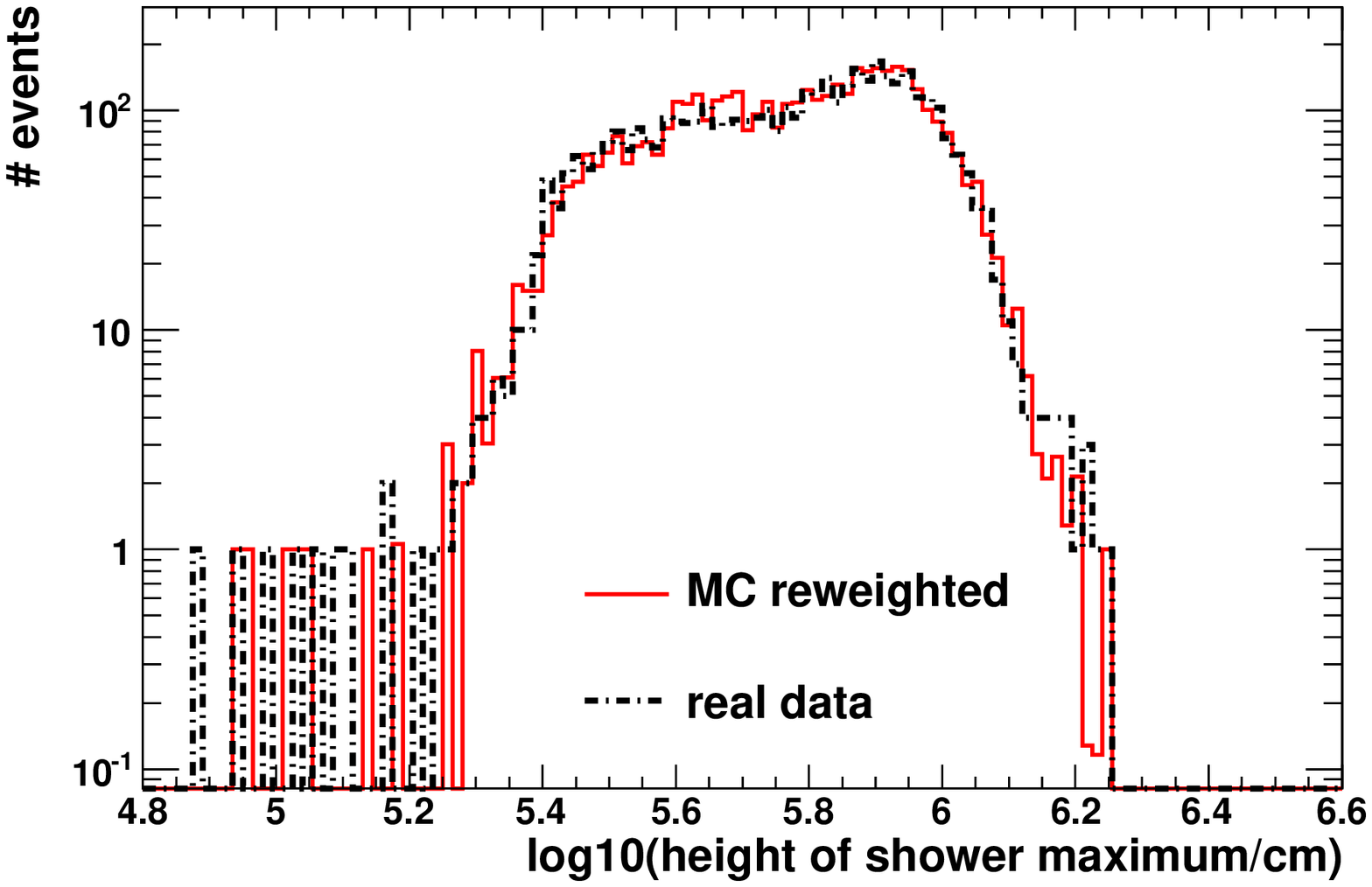}
\vspace*{-1mm}
\end{center}
\end{minipage}
\hspace*{1mm}
\begin{minipage}[b]{0.5\linewidth}
\begin{center}
\includegraphics[width=3.in]{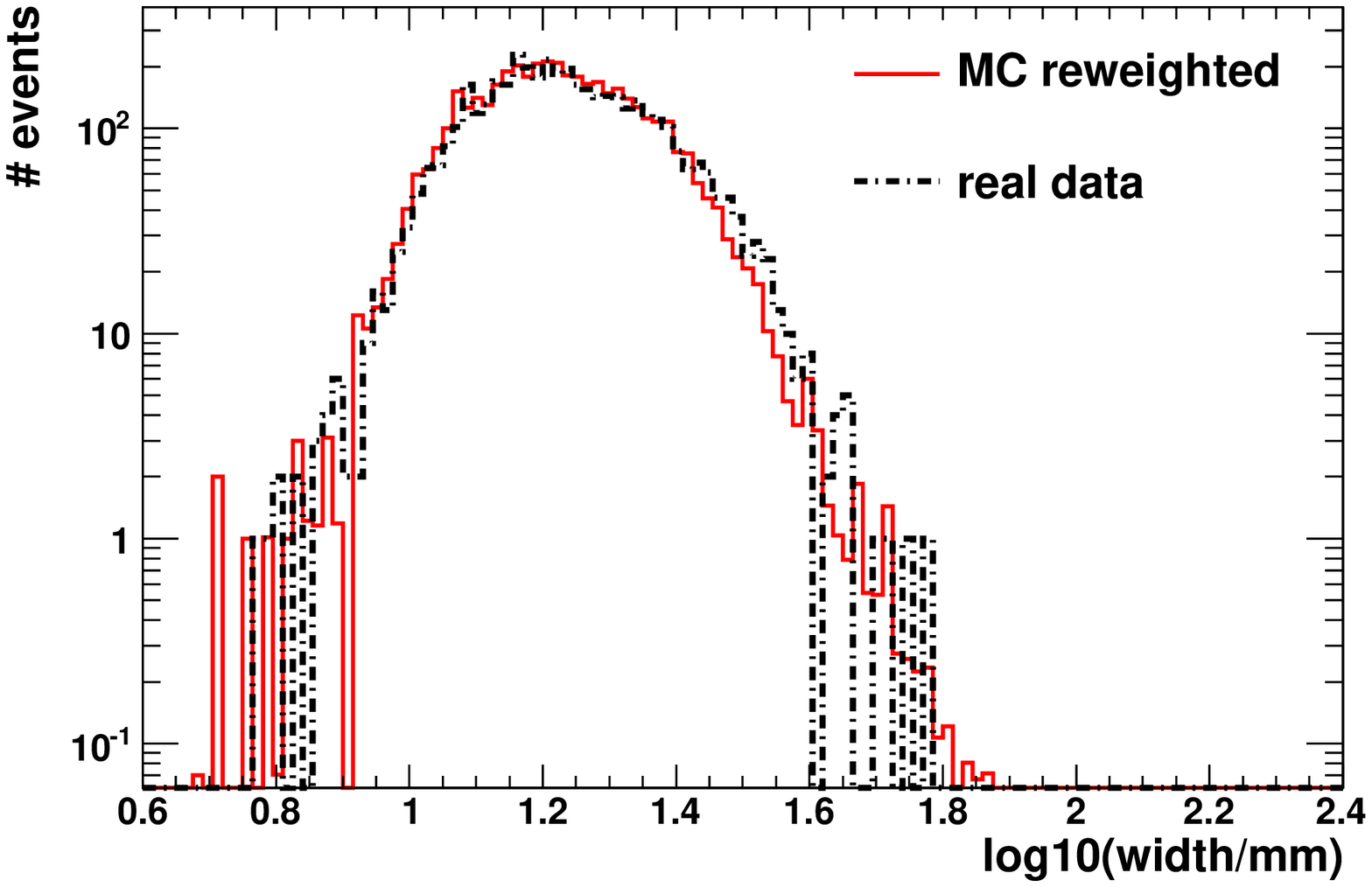}
\vspace*{-1mm}
\end{center}
\end{minipage}
\caption{\label{fig:magiccheckothervariables}Comparison of observable distributions for real data (black/dot-dashed) and re-weighted MC (red/solid). Shown are observables which have not been considered during the unfolding fit, namely \textit{height of shower maximum} and \textit{width}.}
\end{figure*}

The shown result which has been obtained with TRUEE has been verified in terms of the agreement of observable distributions in data and an accordingly weighted MC event sample. Figure~\ref{fig:magiccheckunfoldingvariables} shows the comparison for two observables which have been used during the unfolding, while Fig.~\ref{fig:magiccheckothervariables} displays the distributions for observables which have been neglected during the unfolding. Even for observable parameters which have not been considered during the unfolding process, the a posteriori distributions match very closely. This is a strong confirmation for the quality of the unfolding result.

\subsection{The IceCube neutrino observatory}
\label{sec:icecube}

\subsubsection{Experiment}
IceCube is a cubic kilometer-scale neutrino detector located at the geographic South Pole. The main goal of IceCube is the investigation of cosmic rays by the detection of neutrinos.
Since neutrinos have a very small interaction cross section and thus pass through a large amount of matter, a large detection volume is required to obtain neutrino-induced signals with reasonable statistics. For this reason IceCube utilizes a volume of 1\,km$^3$ in the glacial ice at the depth between 1\,450 and 2\,450\,m, forming a three-dimensional grid of 5\,160 digital optical modules (DOM) which are equipped with photomultipliers. The IceCube DOMs are fixed on strings which are arranged in a triangular pattern in distances of 125\,m to each other. The detector deployment has been executed during antarctic summer seasons, from 2005 till 2011. Each year since then, data has been taken with an $n$-string configuration of the partially constructed detector. For the analysis shown here the IceCube 59 string configuration (IC\,59) is used.

Neutrinos only undergo weak interaction and thus cannot be detected directly. They produce secondary particles, such as muons, electrons or tauons according to the neutrino flavor. These and other secondary charged particles induce Cherenkov light in the ice if their energy is high enough. The Cherenkov light propagates through the ice and causes signals, so-called hits, in the DOMs along the track of the secondary particle within the detector volume. From the time difference and the amount of charge in each DOM, the track of the secondary lepton can be reconstructed and its observable values are saved as one event. Only muons have a track-like signature in the detector and provide sufficient directional information. Therefore, we consider muon neutrinos in the following. 

Muons produced in the Earth's atmosphere represent the main component of the background. In contrast, neutrinos can pass through the Earth, due to their small cross section. Thus, the Earth can be used as a filter to reduce the muon background by considering events coming only from below the horizon.

In this example analysis the interest is focussed on the determination of the flux of muon neutrinos coming from decays of charged pions and kaons that are in turn produced by interactions of cosmic rays with the Earth's atmosphere. Studying the spectrum of this atmospheric neutrino flux at energies beyond $\sim 3\cdot10^{14}$\,eV can provide information about the production of charmed mesons in the atmosphere by showing an enhanced neutrino flux at higher energies, compared to the neutrino flux caused by light meson decays. Furthermore, a flattening of the neutrino energy spectrum to higher energies can permit conclusions about the existence of extragalactic high energy neutrinos, as their predicted flux shows a harder spectral index than the atmospheric neutrino flux. This would reveal new insights concerning the different models of cosmic ray production in the cosmic accelerators, such as Active Galactic Nuclei and Gamma Ray Bursts. Therefore an accurate estimation of the energy spectrum is essential.

\begin{figure*}[!th]
	\begin{minipage}[b]{0.5\linewidth}
	\begin{center}
		\includegraphics[width=3.in]{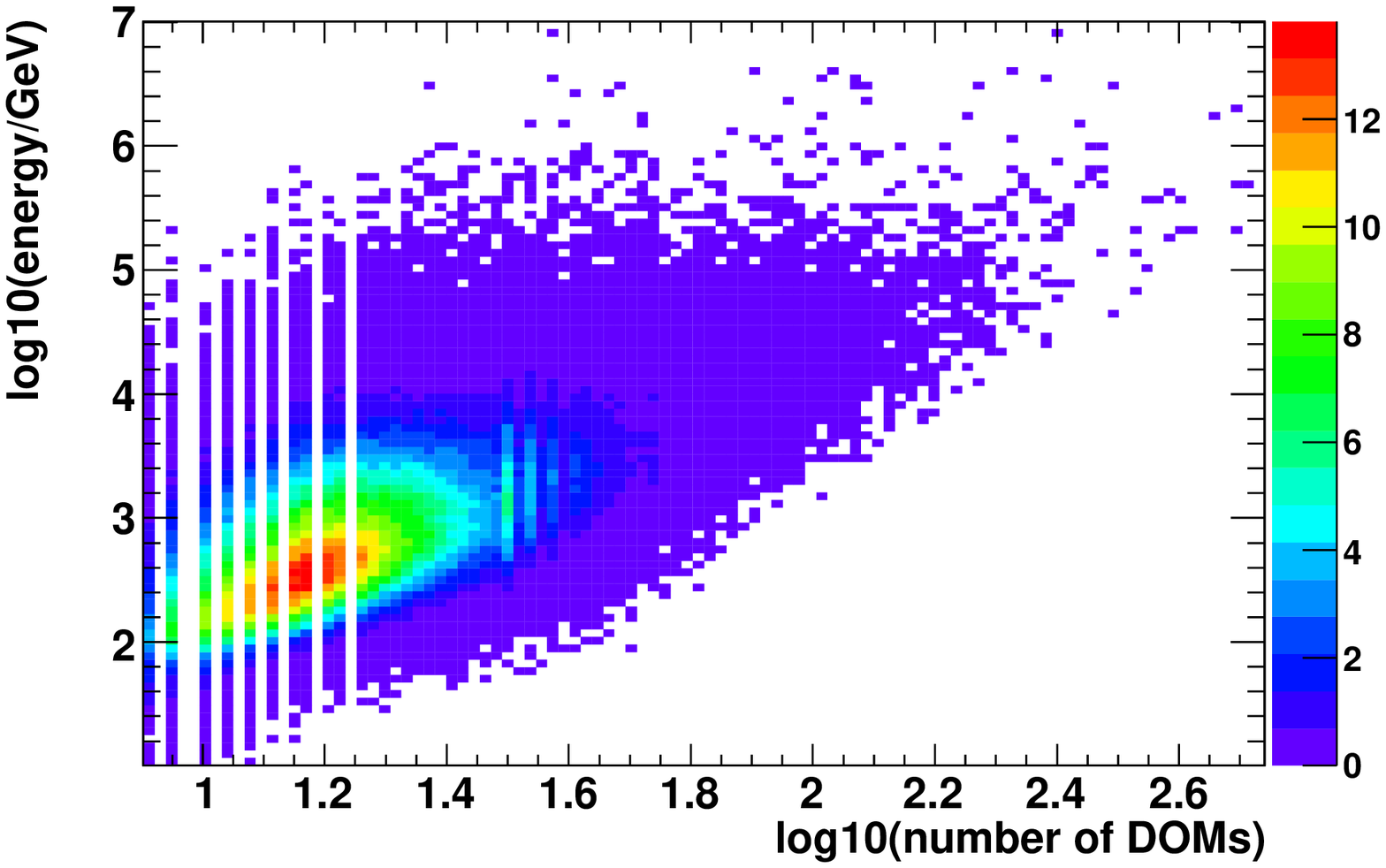}
		\includegraphics[width=3.in]{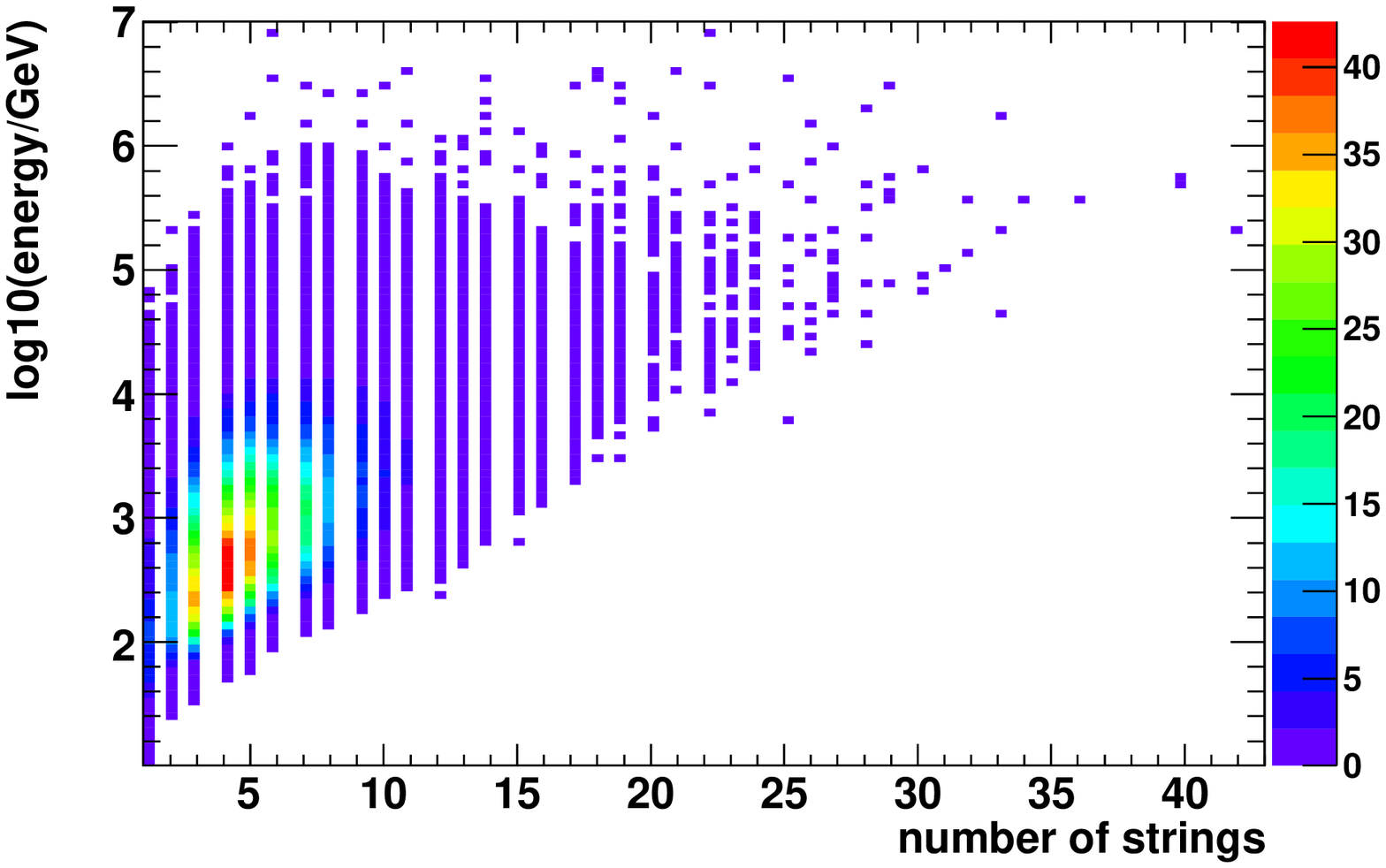}
		\includegraphics[width=3.in]{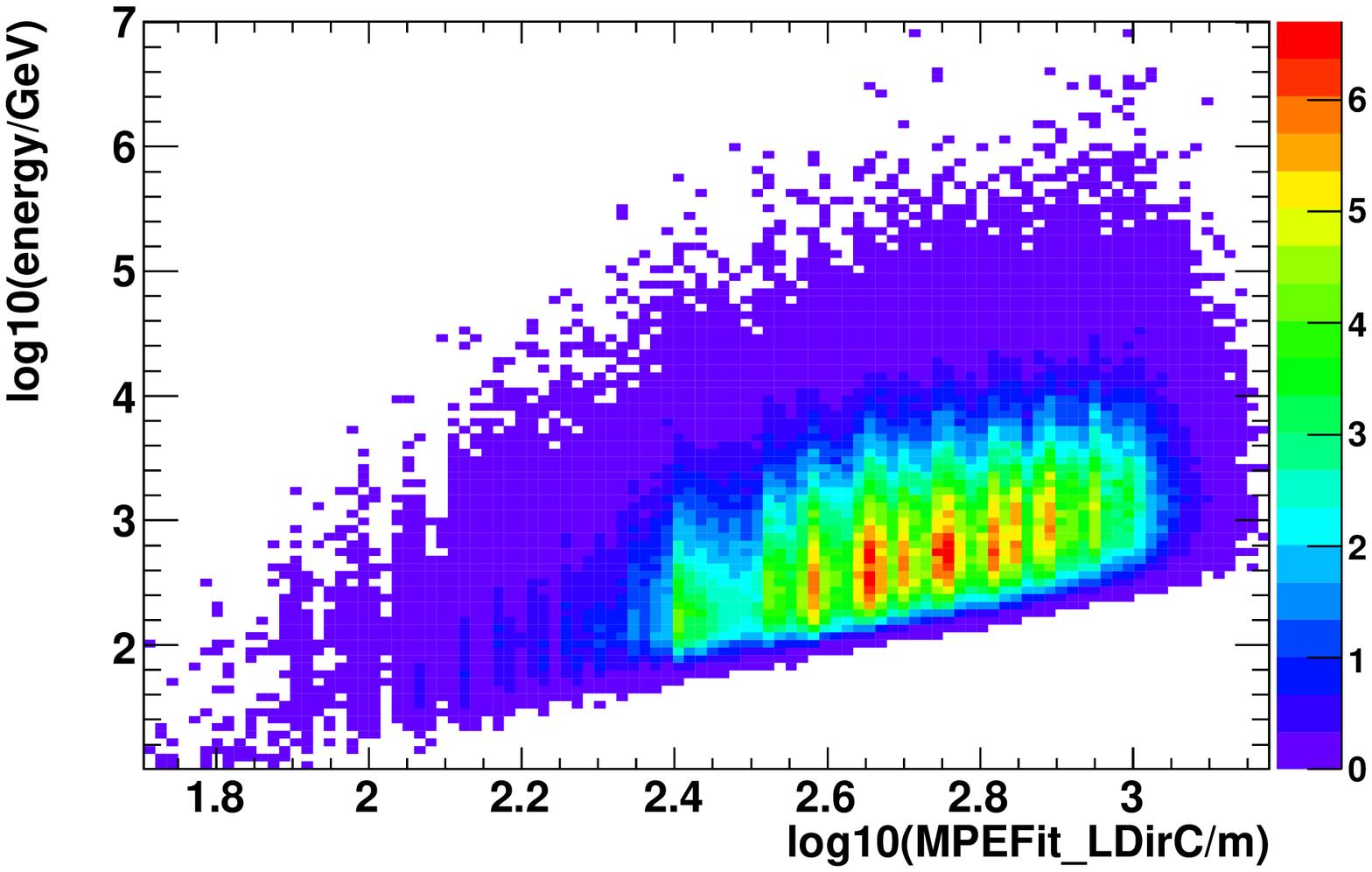}
	\end{center}
	\end{minipage}
	\hspace*{-3mm}
	\begin{minipage}[b]{0.5\linewidth}
	\begin{center}
		\includegraphics[width=3.in]{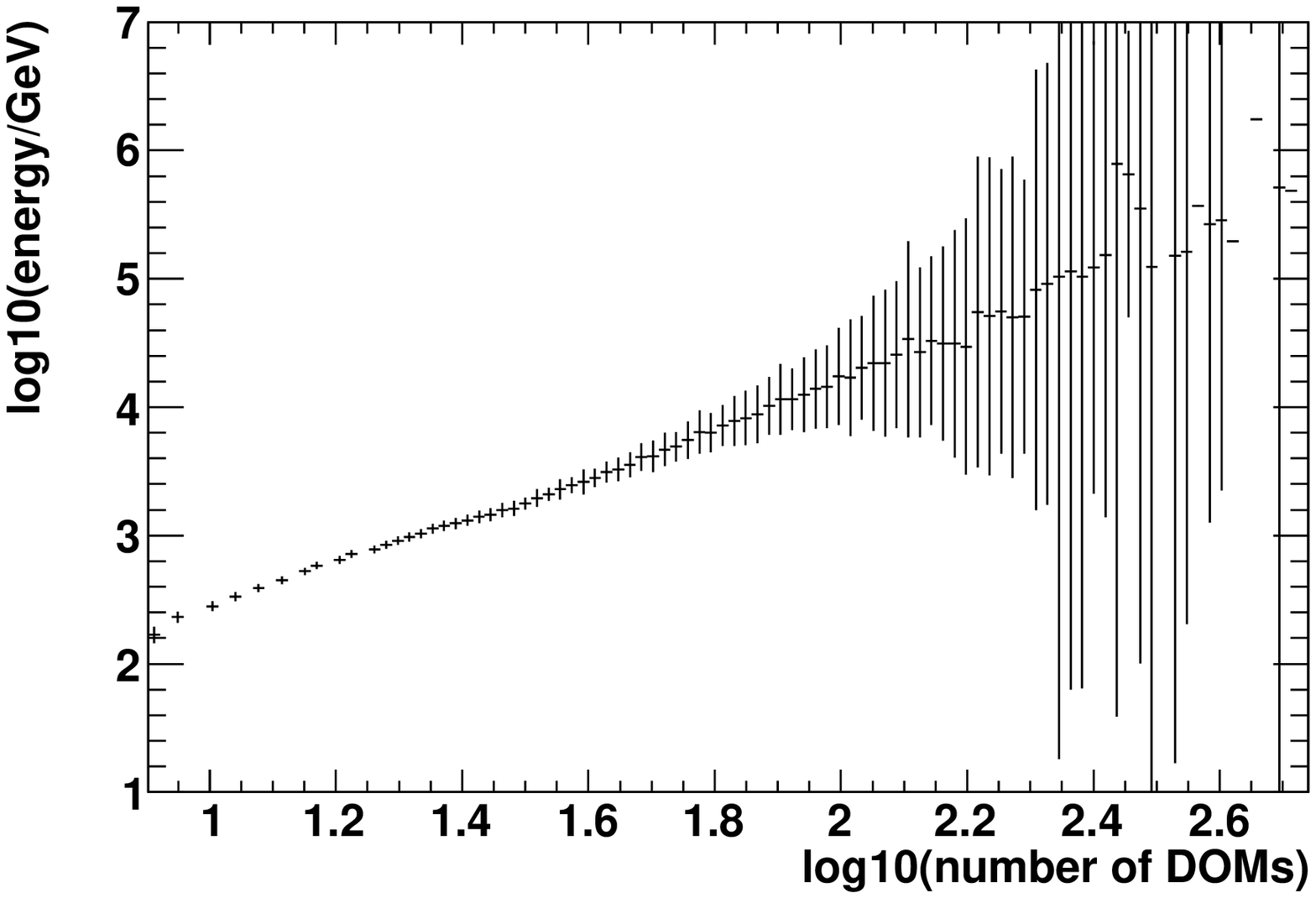}
		\includegraphics[width=3.in]{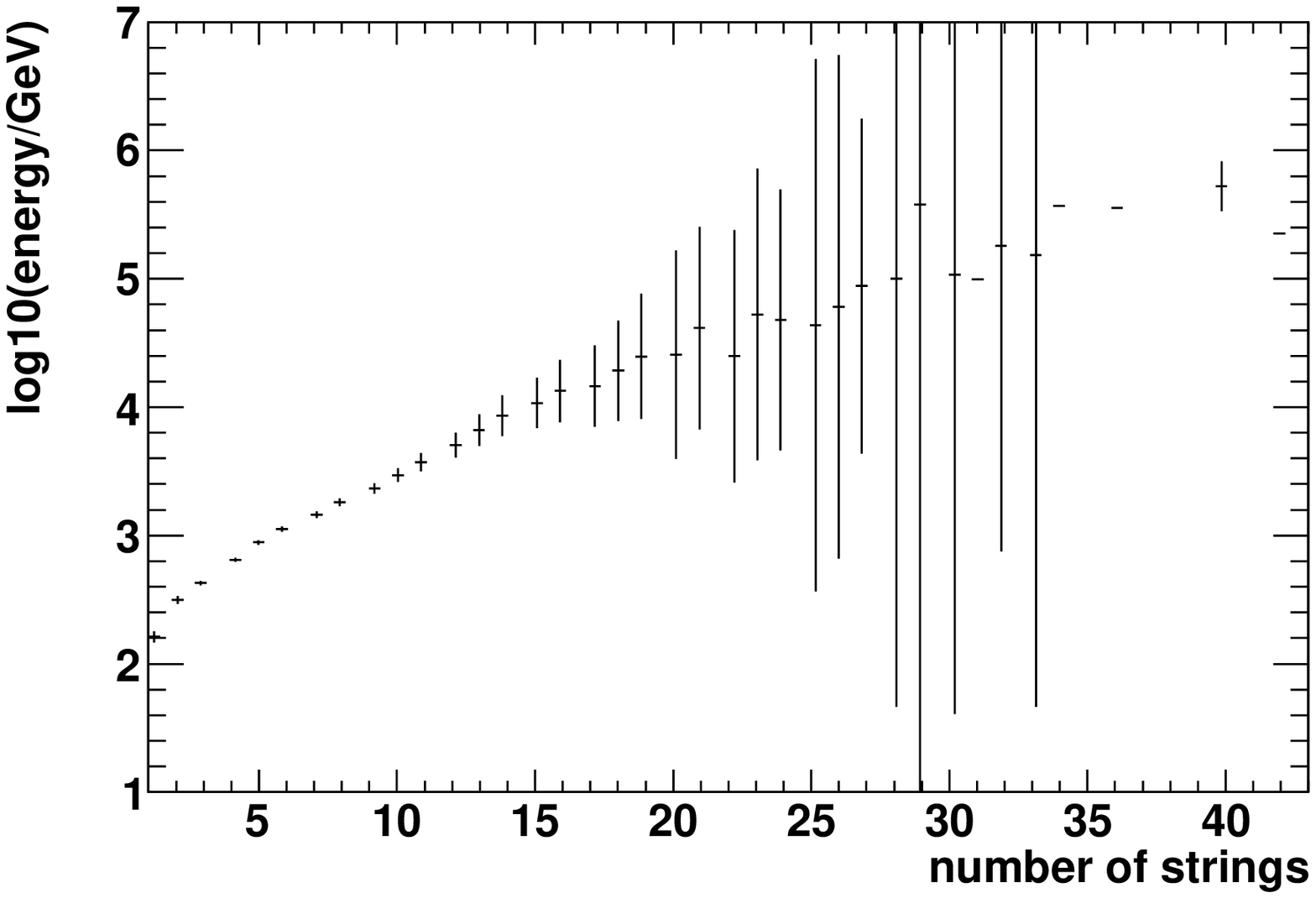}
		\includegraphics[width=3.in]{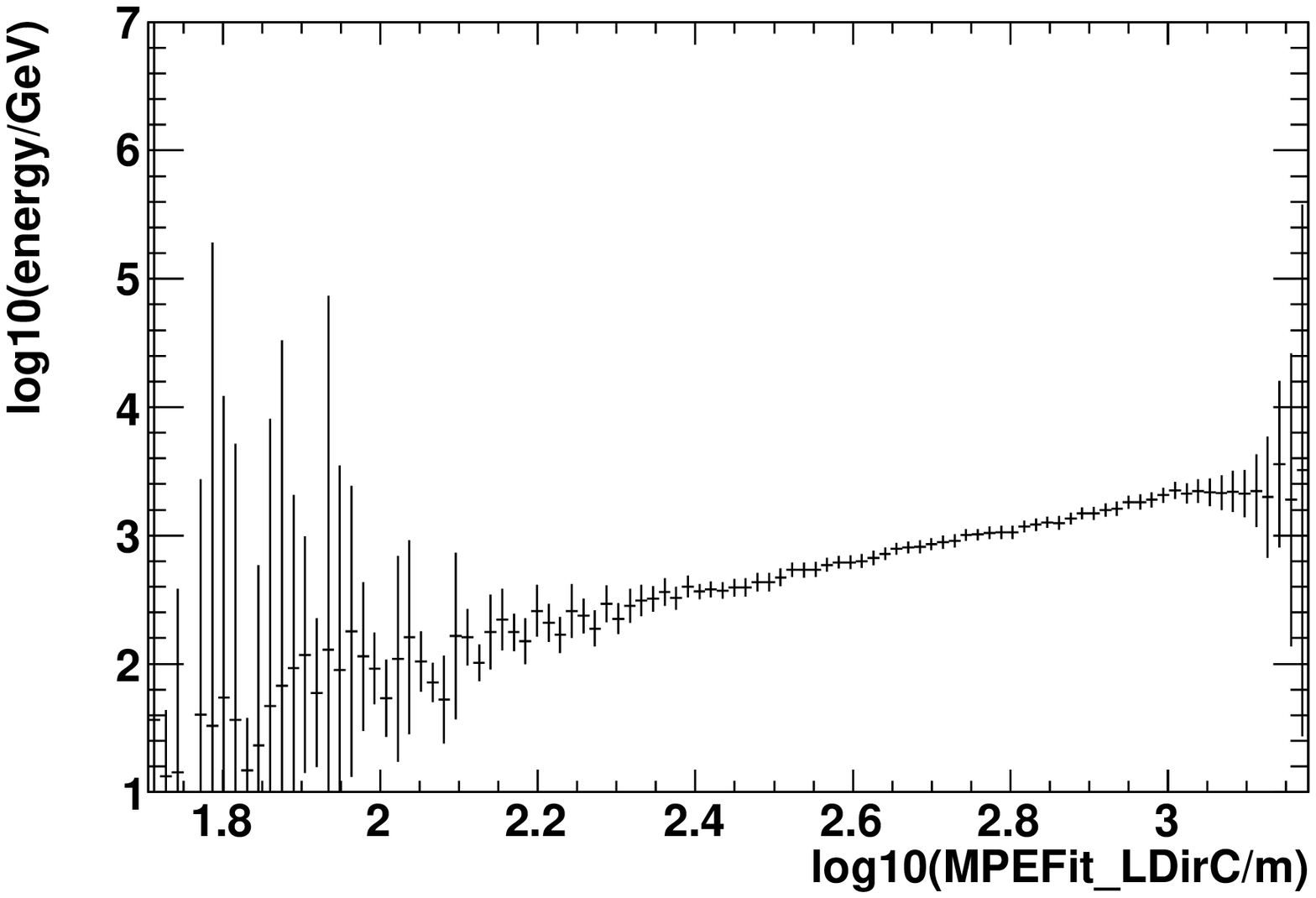}
	\end{center}
	\end{minipage}
		\caption{Scatter plots (left) and related profile histograms (right) used to check the correlation between the energy and the observables.}
	  \label{Fig:correlation_ic}
\end{figure*} 
In the following, the steps of the regularized unfolding of the neutrino energy spectrum with TRUEE are demonstrated, by using Monte Carlo simulations and 10\,\% of the measured IC\,59 data. This analysis serves as a proof of principle and is not supposed to point out any conclusions about neutrino physics.
\subsubsection{Neutrino sample}
For the analysis of the entire neutrino flux a neutrino sample with high purity is desirable. The background contamination caused by mis-reconstructed atmospheric muons is chosen not to exceed 5\,\% to keep the uncertainty of the estimated energy spectrum small compared to statistical uncertainties. In the following, we use a neutrino sample which was obtained in the course of the atmospheric neutrino analysis. To reduce the background and to obtain a neutrino sample with a sufficiently large number of events, series of straight cuts were applied to the data, including the zenith angle cut $\theta =88^{\circ} - 180^{\circ}$. Although the rejection of muon tracks from above the horizon is made, there are still mis-reconstructed background events. Therefore the final event selection was performed using the multivariate method Random Forest in the framework RapidMiner \cite{Fischer:2002p1824}. The final sample consists of $\sim3\,000$ events in the used 10\,\% in the full data sample collected in one year. The corresponding MC sample, which is needed for the event selection training and the further determination of the response matrix during the unfolding, is produced using the simulation of all physical processes following the theoretical models for cross sections and propagations of particles and photons through different kinds of media. The MC neutrino sample consists of more than $6\cdot10^{5}$ events, which are weighted to describe the measured data observables as accurately as possible \cite{Montaruli:2011p1992}. Using event weights, the MC energy spectrum follows the atmospheric neutrino flux with a spectral index $\gamma \sim3.7$ predicted by Honda \cite{Honda:2007p1789}, including the contribution of prompt neutrinos from charm meson decays at higher energies (Naumov \cite{Bugaev:1989p1820}). The simulated background muon sample, which is used to estimate the purity of the data sample, was produced using the air shower simulator CORSIKA.

\subsubsection{Choice of observables}
As a first step, the selection of energy-dependent observables is done. The inspection of the scatter and profile histograms led to the choice of the following three observables.
\begin{itemize}
	\item \textit{Number of DOMs} which show at least one photoelectron. A muon with higher energy induces more Cherenkov light and has a higher track length and thus a higher probability to cause hits in DOMs.
	\item \textit{Number of strings} which contain at least one hit DOM. This observable provides additional directional information since the number of strings is correlated to the zenith angle of the track. Furthermore, the distances between DOMs on the same string are lower than between those on different strings. Thus, this observable is supplemental to the number of DOMs.
	\item The direct track length of the muon in a certain time window (\textit{MPEFit\_LDirC}). The length is calculated by the projection of the number of the direct (not scattered) photons on the reconstructed track as the distance between the two outermost points.
\end{itemize}
The correlation and profile histograms in Fig.~\ref{Fig:correlation_ic} show the dependency between the observables and the energy. In the TRUEE test mode, different binning of the three observables have been tried in order to find the optimal unfolding result with respect to the true distribution.

\subsubsection{Results}
\begin{figure}[!thh]
	\centerline{
		\includegraphics[width=3.in]{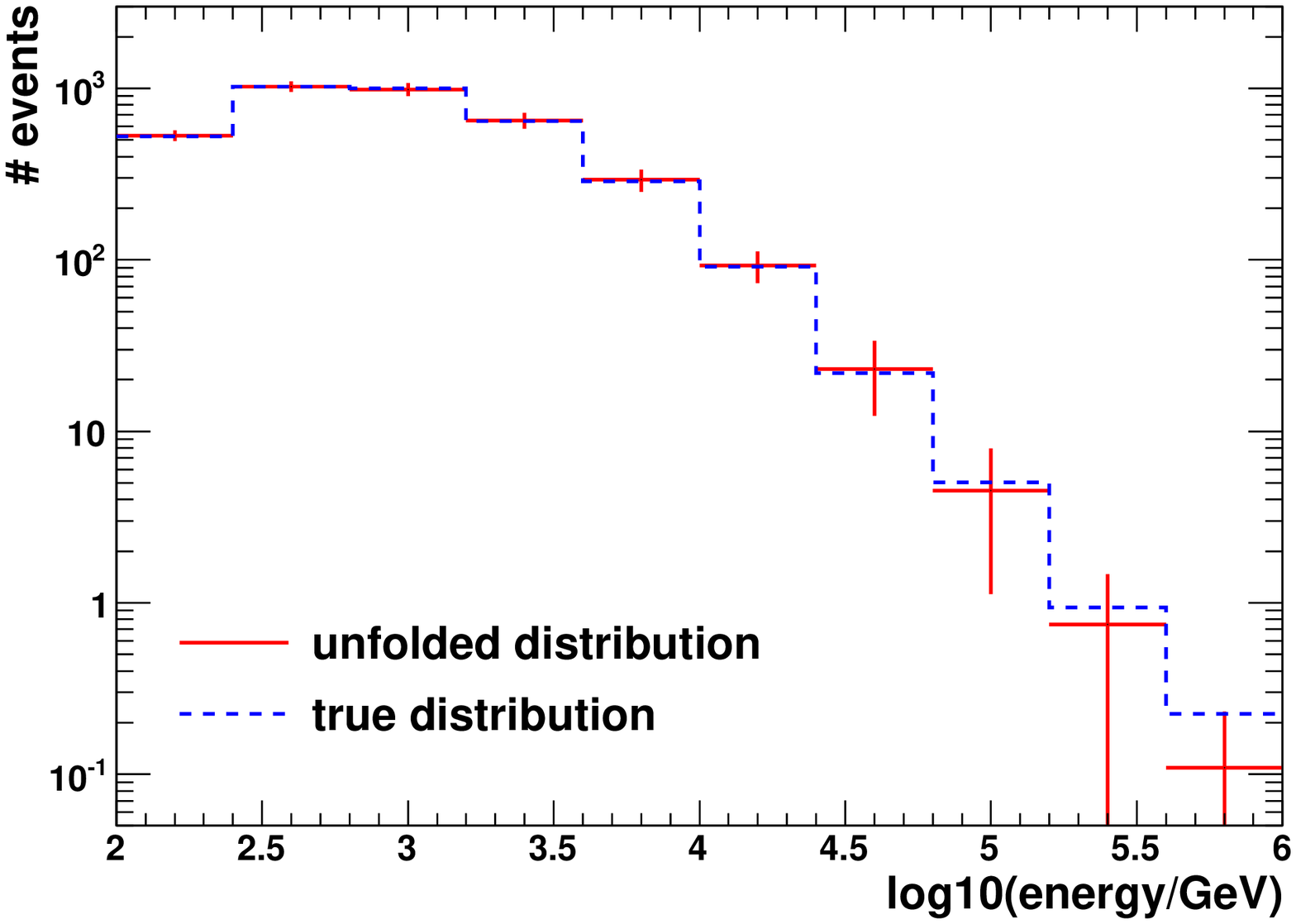}}
	\centerline{
		\includegraphics[width=3.in]{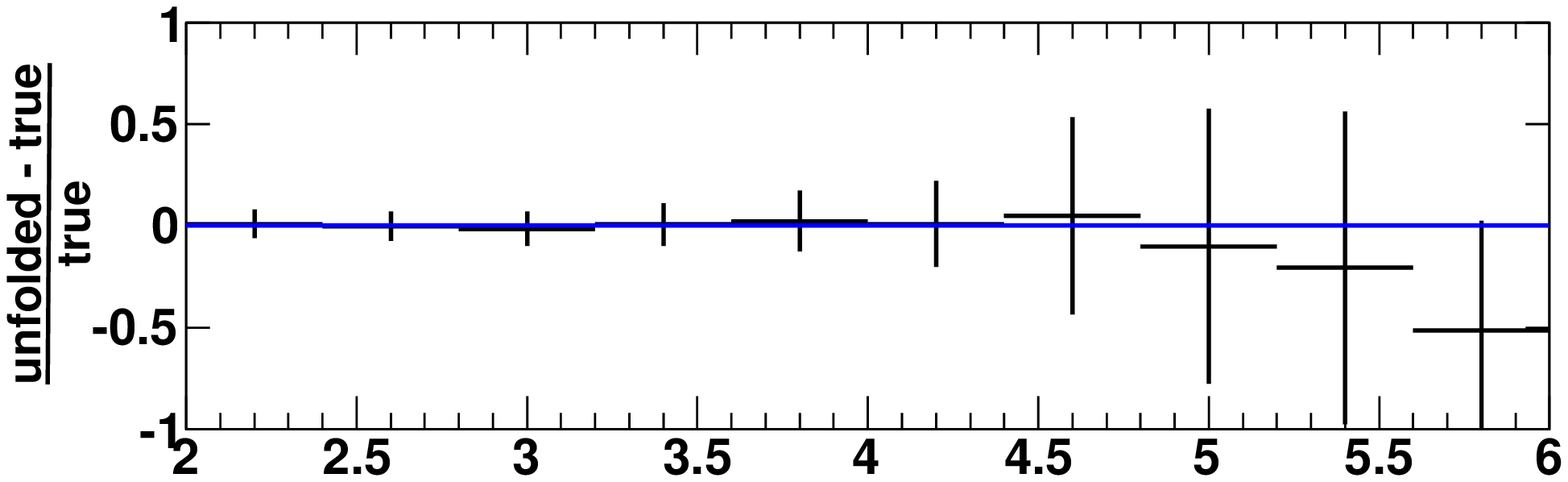}}
  \caption{Test mode result for the final unfolding settings. Shown are the true and the unfolded distribution of the part of the MC sample used for the determination of the response matrix. No acceptance correction is applied. The relative difference between the unfolded and true values is shown in the lower histogram.}
  \label{Fig:test_IC}
\end{figure}

The best result obtained in test mode is shown in Fig.~\ref{Fig:test_IC}. The parameter set which delivered the resulting spectrum consisting of 10 bins is given by 16 knots and 5 degrees of freedom. This combination of parameters is also applied to the real data.

The generated MC neutrino sample which is used to determine the detector response contains only simulated events which undergo an interaction within or close to the detector. This restriction is necessary to reduce simulation time and memory. Therefore the generated function (here following $\sim E^{-2}$) does not consider events, which do not cause any signal in the detector, and cannot be given to the unfolding algorithm to normalize the flux to the correct scale. However, since we use the individual event weights for the MC simulated neutrino sample to make it similar to the real data sample, the reweighted sample follows the atmospheric neutrino flux, calculated by Honda and Naumov. Thus this atmospheric neutrino flux function can be provided to TRUEE to describe the generated neutrino event distribution and make the full acceptance correction. We compare this method to the standard IceCube analysis procedure using the effective area to scale the final result to the original neutrino flux \cite{PhysRevD.83.012001}. This method is described in the following.

\begin{figure}[!thh]
  \centerline{	
	\includegraphics[width=3.in]{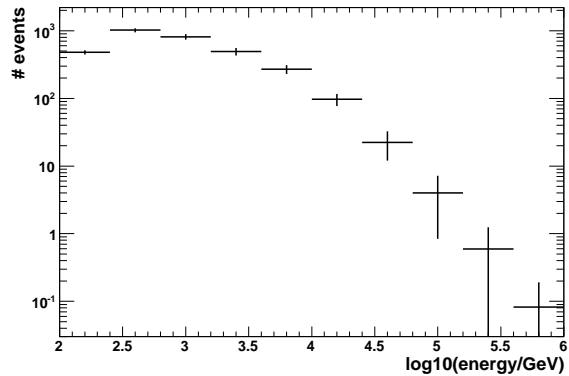}}
  \caption{The unfolding of the IC\,59 neutrino sample gives the distribution of selected neutrino events depending on energy.}
  \label{Fig:unfolded}
\end{figure}

After passing all event selection steps, the final sample contains only a fraction of neutrino events. Thus, the unfolded distribution represents only neutrinos which interacted, triggered the detector and passed the event selection (Fig.~\ref{Fig:unfolded}).

\begin{figure}[!thh]
  \centerline{
		\includegraphics[width=3.in]{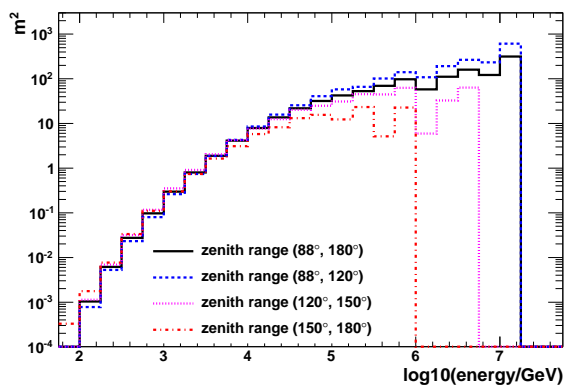}}
  \caption{Effective area for the current neutrino sample dependent on neutrino energy. Illustrated are areas for different zenith angle ranges and for the average of the whole zenith range of $88^{\circ}$ to $180^{\circ}$, which is considered in the analysis.}
  \label{Fig:effArea}
\end{figure}

To calculate the neutrino flux for all neutrinos within the zenith angle range, the unfolded spectrum has to be scaled with the effective area. This is the ratio between the observed event rate and the incoming flux and depends on the properties of the selected event sample and on the energy. It includes the muon neutrino cross section, the probability for the muon to be detected and the detector efficiency for muon detection and event reconstruction. The effective area for the current sample is shown in Fig.~\ref{Fig:effArea}.
It rises at higher energies due to the increasing cross section of neutrinos and the longer tracks of neutrino-induced muons. For the events with vertically upgoing tracks the effective area decreases because of the rising probability for absorption of neutrinos within the Earth.

\begin{figure}[!thh]
  \centerline{
		\includegraphics[width=3.in]{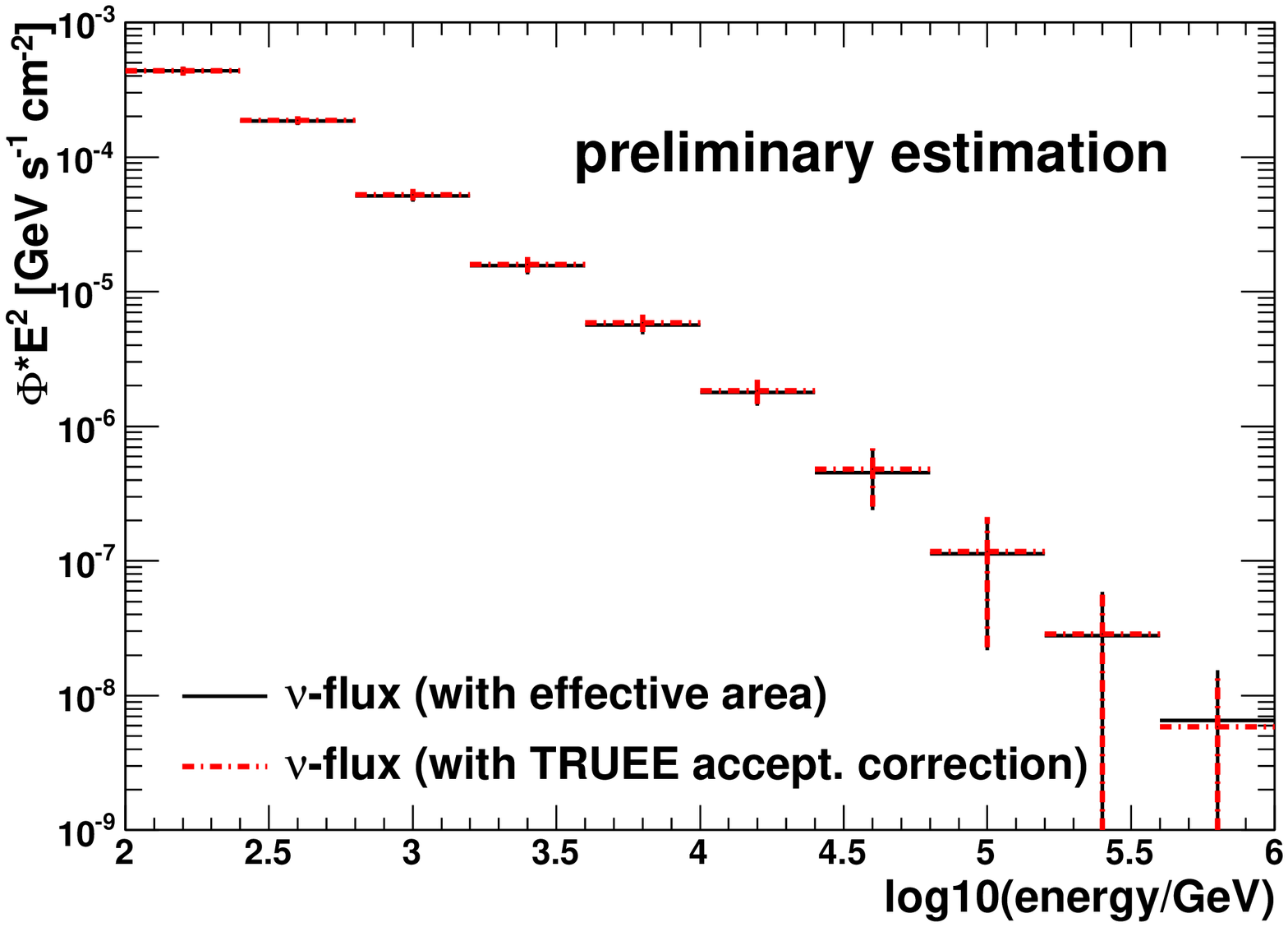}}
	\centerline{
		\includegraphics[width=3.in]{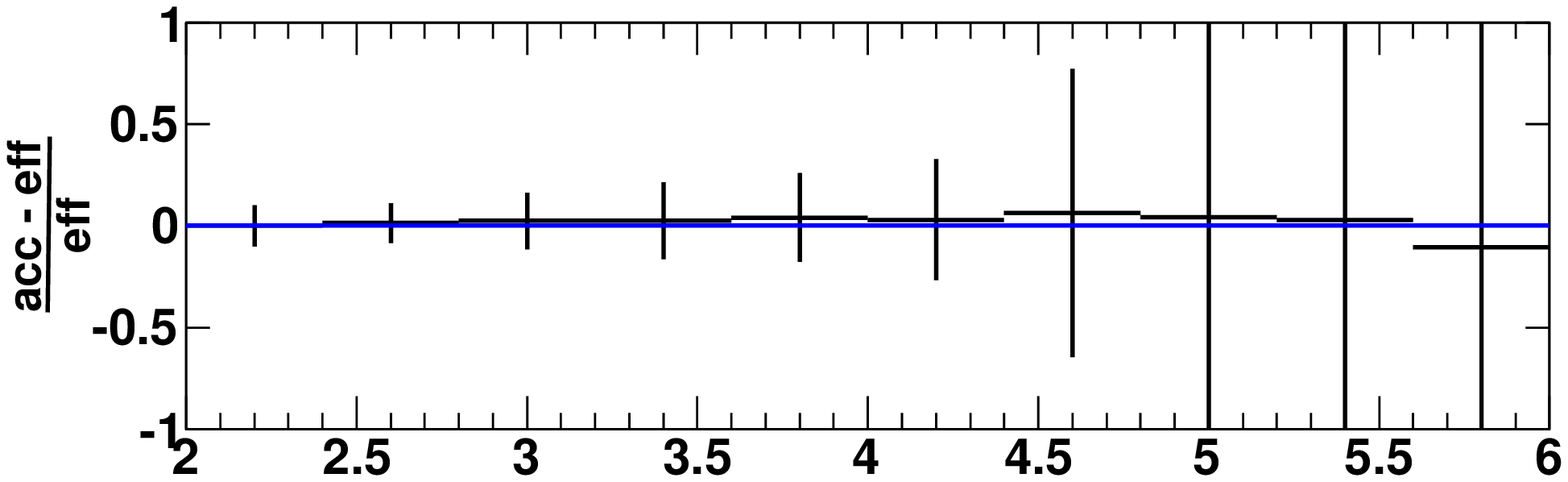}}
  \caption{Examples of an atmospheric neutrino energy spectrum gained from 10\,\% of IC\,59 data unfolded with TRUEE. The two spectra are obtained using different methods of acceptance correction: the standard IceCube method using effective area (black/solid) and the TRUEE internal acceptance correction (red/dot-dashed). The uncertainties are determined by the unfolding software using standard error propagation, while systematics are not considered in these results. The spectra are weighted with the square of the energy. The relative difference between both distributions is demonstrated in the lower histogram.}
  \label{Fig:result_IC}
\end{figure}

In Fig.~\ref{Fig:result_IC}, an example of a neutrino flux spectrum is shown, which can be derived from an unfolded energy distribution of neutrino events (Fig.~\ref{Fig:unfolded}), if the effective area (Fig.~\ref{Fig:effArea}) is known. Additionally, we present the result which has been obtained with the internal acceptance correction by providing the function of the atmospheric neutrino flux to TRUEE (see also Fig.~\ref{Fig:result_IC}).

\subsubsection{Verification}
\label{sec:verificationIcecube}
\begin{figure}[!thh]
	\centerline{
	  \includegraphics[width=3.in]{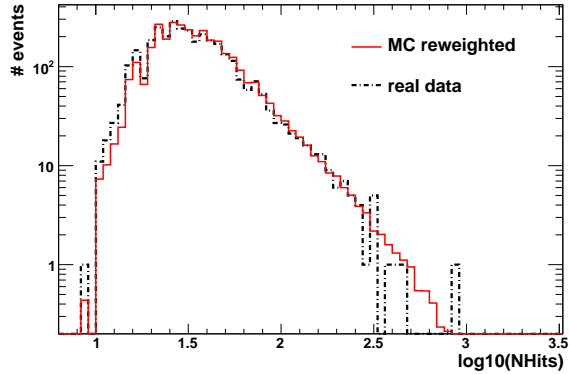}}
	\caption{Comparison of data (black/dot-dashed) and MC (red/solid) weighted to the unfolded function. Shown is the number of hits detected in all DOMs per event.}
\label{Fig:verification_IC_1}
\end{figure}

\begin{figure}[!thh]
			\centerline{
	 			\includegraphics[width=3.in]{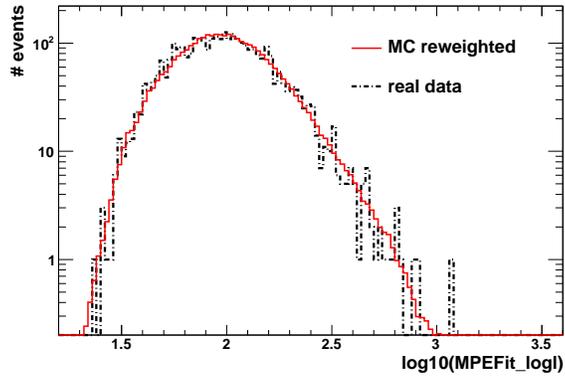}}
				\caption{Comparison of data (black/dot-dashed) and MC (red/solid) weighted to the unfolded function. Shown is the log-Likelihood value of an event reconstruction fit.}
  \label{Fig:verification_IC_2}
\end{figure}

 To check the quality of the unfolding the agreement between the real data and the weighted Monte Carlo sample is investigated. Verification histograms of two observables which have not been used for the unfolding fit are shown in Fig.~\ref{Fig:verification_IC_1} and Fig.~\ref{Fig:verification_IC_2}.

\subsection{Tests on the influence of the simulation}
Generally, the unfolding permits the estimation of an unknown distribution. With the determined response matrix, the unfolding should be able to identify any distribution in the data, independent from the distribution of the simulations which have been used for the determination of the response matrix. The only requirements are that the simulation model describes the data well enough and that all bins of the observables are filled with a sufficient number of MC events. In general a ten times larger number of simulated events compared to data events is enough to neglect the uncertainties in the response matrix \cite{Blobel:2002p1184}.

In some cases the unfolded distribution can have a very steep spectral slope. This is also true for astroparticle problems. To ensure that enough MC events are contained in the high energy region, the spectral slope of the simulated distribution should not deviate too much from the true data distribution. The following tests are made to investigate the impact of the deviation in the spectral distributions between the MC sample and the pseudo data sample by using different spectral slopes in the simulation. We use different toy MC samples for the calculation of the response matrix and unfolding which describe the following distributions with respect to the arbitrary variable $x$

\begin{figure}[!thh]
	\centerline{\includegraphics[width=0.50\textwidth]{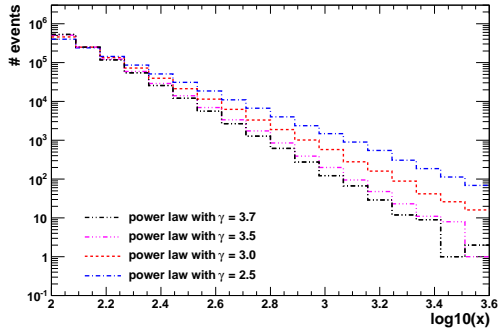}}
	\caption{Four toy MC samples used for the determination of the individual response matrices. The simulated distributions have different spectral slopes.}
	\label{Fig:slope_influence_MC}
\end{figure}

\begin{itemize}
	\item power law with $\gamma = 3.7$ (related to the atmospheric neutrino flux)
	\item power law with $\gamma = 3.5$
	\item power law with $\gamma = 3.0$
	\item power law with $\gamma = 2.5$.
\end{itemize}
The different MC simulations which are used for the determination of the response matrix are shown in Fig.~\ref{Fig:slope_influence_MC}.

\begin{figure}[!thh]
	\centerline{\includegraphics[width=0.50\textwidth]{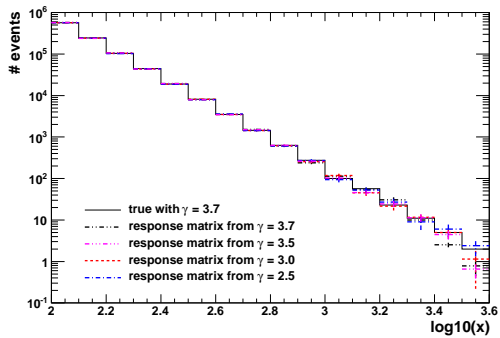}}
	\caption{Unfolding results of the simulated pseudo data sample following the power law with $\gamma = 3.7$ using different MC simulated distributions for the response matrix calculation. The true sought distribution is shown by the solid line.}
	\label{Fig:slope_influence_37}
\end{figure}

\begin{figure}[!thh]
	\centerline{\includegraphics[width=0.50\textwidth]{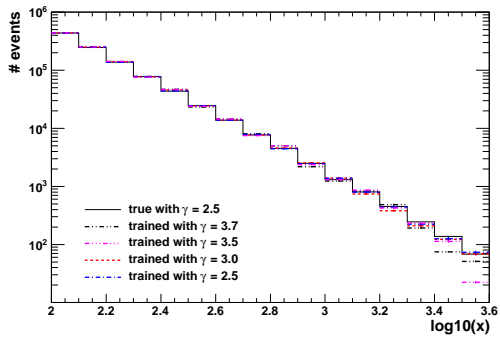}}
	\caption{Unfolding results of the simulated pseudo data sample following the power law with $\gamma = 2.5$ using different MC simulated distributions for the response matrix calculation. The true sought distribution is shown by the solid line.}
	\label{Fig:slope_influence_25}
\end{figure}

We present unfolding results of two pseudo data samples with the steepest ($\gamma = 3.7$) and the flattest ($\gamma = 2.5$) power law.
The results are shown in Fig.~\ref{Fig:slope_influence_37} and Fig.~\ref{Fig:slope_influence_25}.
The slope of the pseudo data sample is the same or steeper than the slope of the MC samples. The maximum deviation between the spectral indices of MC and pseudo-data is 1.2. The unfolding results with the flatter MC assumptions are consistent with the true distribution within the uncertainties. The MC sample with the steepest spectral slope causes an underestimation of the event distribution at x-values greater than $log_{10}(x)=3.4$. This is caused by the fact that the response matrix is not well enough described due to the low amount of events. Instead, the x-region below $log_{10}(x)=3.4$ has bins containing more than 10 events and thus a good agreement of the unfolded result with the true distribution is ensured. The same is true for the unfolding of the power law distribution with $\gamma = 2.5$.

The conclusion of the test is the recommendation to use a MC sample for the response matrix which features a similar or harder spectrum compared to the real data, especially if the unfolded distribution covers several orders of magnitude. The bins of the MC sought distribution should contain at least 10 events. In case of a completely unknown true distribution, an iterative approach of matching the MC spectral slope to the real data can be executed.

\section{Summary and outlook}

The new unfolding software TRUEE has been tested within the astroparticle experiments MAGIC and IceCube and appears to be a very suitable tool for astroparticle physics, since it can properly estimate distributions which cover several orders of magnitude.
The input of all samples is event-wise, thus the response matrix is calculated with an individual binning for every distinct case. A moderate deviation of the distribution of simulated events, used to determine the response matrix, from the data distribution is tolerable, thus an a-priori knowledge of the exact spectral slope of the estimated distribution is not necessary.

In TRUEE, the uncertainties are calculated in the same way as was done in ${\cal RUN}$. They have been proven to follow the Poisson distribution or, in the case of a large number of events, the Gaussian distribution~\cite{Muenich:2007p1821}. Therefore the exclusion of values outside the uncertainties can be made with the minimal probability of 68\,\%. An additional investigation to calculate confidence intervals is being developed within the collaborative research center SFB~823. Within the same project the time dependency of TRUEE will be implemented. Instead of time-slices, the unfolding will be able to deliver a two-dimensional distribution. This is suggested by the fact that a simple fragmentation of the data sample in several packages along the time axis is unacceptable to get reasonable results in cases of low statistics.

Furthermore, an option which allows to automatically perform a second unfolding iteration with a re-weighted MC sample will be implemented. Thereby potentially large differences in the spectral distributions in MC and data can be avoided, which is especially important for the case of an acceptance correction within TRUEE. While the procedure itself has been presented and verified here, only its integration into the program is still to be performed.

TRUEE has been developed in the programming language C++. It contains the proven algorithm ${\cal RUN}$ and additional user-friendly functions, which offer a more comfortable handling of an unfolding analysis. The software is easy to install and convenient to use in combination with modern software. TRUEE and the original algorithm ${\cal RUN}$ deliver comparable results.

TRUEE is intended to be included in the common framework for unfolding software RooUnfold \cite{Adye:2011p1825}. Additionally, TRUEE is currently tested by several particle physics groups.

Within the collaborative research center SFB~823, the fields of application of the program TRUEE will be expanded to solving problems in the context of economics and engineering.

The TRUEE software and the user manual can be found at http://app.tu-dortmund.de/TRUEE/.
\section*{Acknowledgment}
This research is supported by the DFG through the Collaborative Research Center SFB~823, project C\,4 and the DAAD PPP-Spanien program, which are gratefully acknowledged.
We acknowledge the support from the collaborations of the experiments MAGIC and IceCube.
Special thanks are directed to the members of the ``Statistics Tools Group'' of the Helmholtz Alliance ``Physics At The Terascale'' for productive discussions. We would also like to thank Julian Krause for helpful discussions on the MAGIC data analysis.
\bibliographystyle{model1-num-names}
\bibliography{Bibliographie}

\begin{thebibliography}{30}
\expandafter\ifx\csname natexlab\endcsname\relax\def\natexlab#1{#1}\fi
\providecommand{\bibinfo}[2]{#2}
\ifx\xfnm\relax \def\xfnm[#1]{\unskip,\space#1}\fi
\bibitem[{Blobel(1985)}]{Blobel:1984p1896}
\bibinfo{author}{V.~Blobel},
\newblock \bibinfo{title}{Unfolding methods in high energy physics
  experiments},
\newblock \bibinfo{journal}{CERN Yellow Report 85-02}  (\bibinfo{year}{1985}).
\bibitem[{{Fredholm}(1903)}]{fredholm}
\bibinfo{author}{E.~{Fredholm}},
\newblock \bibinfo{title}{{Sur une classe d'{\'e}quations fonctionnelles}},
\newblock \bibinfo{journal}{Acta Math.} \bibinfo{volume}{27}
  (\bibinfo{year}{1903}) \bibinfo{pages}{365--390}.
\bibitem[{de~Boor(2001)}]{deBoor:2001p1990}
\bibinfo{author}{C.~de~Boor}, \bibinfo{title}{{A Practical Guide to Splines}},
  \bibinfo{publisher}{Springer}, \bibinfo{year}{2001}.
\bibitem[{Tikhonov(1963)}]{tikhonov1977solutions}
\bibinfo{author}{A.~Tikhonov},
\newblock \bibinfo{title}{On the solution of ill-posed problems and the method
  of regularization},
\newblock \bibinfo{journal}{Dokl. Akad. Nauk SSSR} \bibinfo{volume}{151}
  (\bibinfo{year}{1963}) \bibinfo{pages}{501 -- 504}.
\bibitem[{Blobel and Lohrmann(1998)}]{Blobel:1998p1987}
\bibinfo{author}{V.~Blobel}, \bibinfo{author}{E.~Lohrmann},
  \bibinfo{title}{{Statistische und numerische Methoden der Datenanalyse}},
  \bibinfo{publisher}{Teubner}, \bibinfo{year}{1998}.
\bibitem[{Blobel(1996)}]{Blobel:1996p1382}
\bibinfo{author}{V.~Blobel},
\newblock \bibinfo{title}{{{The RUN manual: regularized unfolding for
  high-energy physics experiments.}}},
\newblock \bibinfo{journal}{{OPAL Technical Note TN361}}
  (\bibinfo{year}{1996}).
\bibitem[{Martin and Hoffman(2003)}]{Martin:2003p1633}
\bibinfo{author}{K.~Martin}, \bibinfo{author}{W.~Hoffman},
\newblock \bibinfo{title}{{The CMake Build Manager - Cross platform and open
  source}},
\newblock \bibinfo{journal}{Dr. Dobbs}  (\bibinfo{year}{2003}).
\bibitem[{{Chakravarti} et~al.(1967){Chakravarti}, {Laha}, and
  {Roy}}]{Chakravarti:1967}
\bibinfo{author}{I.~{Chakravarti}}, \bibinfo{author}{R.~{Laha}},
  \bibinfo{author}{J.~{Roy}}, \bibinfo{title}{{{Handbook of Methods of Applied
  Statistics}}}, volume~\bibinfo{volume}{1}, \bibinfo{publisher}{{John Wiley
  and Sons}}, \bibinfo{year}{1967}.
\bibitem[{Carmona et~al.(2011)}]{performancepaper}
\bibinfo{author}{E.~Carmona}, et~al.,
\newblock \bibinfo{title}{{Performance of the MAGIC Stereo System}},
\newblock \bibinfo{journal}{Proc. 32nd ICRC, Beijing, China}
  (\bibinfo{year}{2011}).
\bibitem[{Halzen and Klein(2010)}]{Halzen:2010p1795}
\bibinfo{author}{F.~Halzen}, \bibinfo{author}{S.~R. Klein},
\newblock \bibinfo{title}{{IceCube: An Instrument for Neutrino Astronomy}},
\newblock \bibinfo{journal}{Rev.Sci.Instrum.} \bibinfo{volume}{81}
  (\bibinfo{year}{2010}) \bibinfo{pages}{081101}.
\bibitem[{{Bock} et~al.(2004)}]{Bock:multivariate}
\bibinfo{author}{R.~K. {Bock}}, et~al.,
\newblock \bibinfo{title}{{Methods for multidimensional event classification: a
  case study using images from a Cherenkov gamma-ray telescope}},
\newblock \bibinfo{journal}{Nucl.Instrum.Meth.} \bibinfo{volume}{A516}
  (\bibinfo{year}{2004}) \bibinfo{pages}{511--528}.
\bibitem[{{Voigt} et~al.(2011)}]{Voigt:cut-own}
\bibinfo{author}{T.~{Voigt}}, et~al.,
\newblock \bibinfo{title}{{{Threshold Optimization for Classification in
  Imbalanced Data with Unknown Misclassification Costs}}}
  (\bibinfo{year}{2011}). \bibinfo{note}{Submitted to Advances in Data Analysis
  and Classification}.
\bibitem[{{Albert} et~al.(2008)}]{magic:rf}
\bibinfo{author}{J.~{Albert}}, et~al.,
\newblock \bibinfo{title}{{Implementation of the Random Forest Method for the
  Imaging Atmospheric Cherenkov Telescope MAGIC}},
\newblock \bibinfo{journal}{Nucl.Instrum.Meth.} \bibinfo{volume}{A588}
  (\bibinfo{year}{2008}) \bibinfo{pages}{424--432}.
\bibitem[{{Fomin} et~al.(1994)}]{wobble}
\bibinfo{author}{V.~P. {Fomin}}, et~al.,
\newblock \bibinfo{title}{{New methods of atmospheric Cherenkov imaging for
  gamma-ray astronomy. I. The false source method}},
\newblock \bibinfo{journal}{Astroparticle Physics} \bibinfo{volume}{2}
  (\bibinfo{year}{1994}) \bibinfo{pages}{137--150}.
\bibitem[{Moralejo et~al.(2009)}]{Moralejo:2009p1858}
\bibinfo{author}{A.~Moralejo}, et~al.,
\newblock \bibinfo{title}{{MARS, the MAGIC Analysis and Reconstruction
  Software}},
\newblock \bibinfo{journal}{arXiv:0907.0943}  (\bibinfo{year}{2009}).
\bibitem[{{Hillas}(1985)}]{hillas}
\bibinfo{author}{A.~M. {Hillas}},
\newblock \bibinfo{title}{{Cerenkov light images of EAS produced by primary
  gamma}},
\newblock \bibinfo{journal}{Proc. 19th ICRC, San Diego, California}
  (\bibinfo{year}{1985}) \bibinfo{pages}{445--448}.
\bibitem[{Breiman(2001)}]{Breiman:2001p1823}
\bibinfo{author}{L.~Breiman},
\newblock \bibinfo{title}{{{Random Forests}}},
\newblock \bibinfo{journal}{Machine Learning} \bibinfo{volume}{45}
  (\bibinfo{year}{2001}) \bibinfo{pages}{5--32}.
\bibitem[{Heck et~al.(1998)}]{1998cmcc.book.....H}
\bibinfo{author}{D.~Heck}, et~al.,
\newblock \bibinfo{title}{{CORSIKA: A Monte Carlo code to simulate extensive
  air showers}},
\newblock \bibinfo{journal}{Wissenschaftliche Berichte FZKA 6019}
  (\bibinfo{year}{1998}).
\bibitem[{Carmona et~al.(2007)}]{magicmc}
\bibinfo{author}{E.~Carmona}, et~al.,
\newblock \bibinfo{title}{{Monte Carlo Simulation for the MAGIC-II System}},
\newblock \bibinfo{journal}{Proc. 30th ICRC, Merida, Yucatan, Mexico}
  (\bibinfo{year}{2007}) \bibinfo{pages}{1373--1376}.
\bibitem[{Albert et~al.(2007)}]{Collaboration:2007p1826}
\bibinfo{author}{J.~Albert}, et~al.,
\newblock \bibinfo{title}{{Unfolding of differential energy spectra in the
  MAGIC experiment}},
\newblock \bibinfo{journal}{Nucl.Instrum.Meth.} \bibinfo{volume}{A583}
  (\bibinfo{year}{2007}) \bibinfo{pages}{494--506}.
\bibitem[{Bertero(1989)}]{Bertero:1988eh}
\bibinfo{author}{M.~Bertero},
\newblock \bibinfo{title}{{Linear inverse and ill posed problems}},
\newblock \bibinfo{journal}{Advances in Electronics and Electron Physics, ed.
  P.W.Hawkes} \bibinfo{volume}{75} (\bibinfo{year}{1989})
  \bibinfo{pages}{1--120}. \bibinfo{note}{INFN/TC-88/2}.
\bibitem[{Schmelling(1994)}]{schmelling}
\bibinfo{author}{M.~Schmelling},
\newblock \bibinfo{title}{{The Method of reduced cross entropy: A General
  approach to unfold probability distributions}},
\newblock \bibinfo{journal}{Nucl.Instrum.Meth.} \bibinfo{volume}{A340}
  (\bibinfo{year}{1994}) \bibinfo{pages}{400--412}.
\bibitem[{Fischer et~al.(2002)}]{Fischer:2002p1824}
\bibinfo{author}{S.~Fischer}, et~al., \bibinfo{title}{{{Yale: Yet Another
  Learning Environment}}}, \bibinfo{type}{Technical Report}
  \bibinfo{number}{CI-136/02}, Collaborative Research Center 531, University of
  Dortmund, \bibinfo{year}{2002}. \bibinfo{note}{ISSN 1433-3325.
  {http://yale.sf.net/ }}.
\bibitem[{Montaruli and Ronga(2011)}]{Montaruli:2011p1992}
\bibinfo{author}{T.~Montaruli}, \bibinfo{author}{F.~Ronga},
\newblock \bibinfo{title}{{Comparison of muon and neutrino times from decays of
  mesons in the atmosphere}},
\newblock \bibinfo{journal}{arXiv:1109.6238}  (\bibinfo{year}{2011}).
\bibitem[{Honda et~al.(2007)}]{Honda:2007p1789}
\bibinfo{author}{M.~Honda}, et~al.,
\newblock \bibinfo{title}{{Calculation of atmospheric neutrino flux using the
  interaction model calibrated with atmospheric muon data}},
\newblock \bibinfo{journal}{Phys.Rev.} \bibinfo{volume}{D75}
  (\bibinfo{year}{2007}) \bibinfo{pages}{043006}.
\bibitem[{Bugaev et~al.(1989)}]{Bugaev:1989p1820}
\bibinfo{author}{E.~Bugaev}, et~al.,
\newblock \bibinfo{title}{{Prompt Leptons in Cosmic Rays}},
\newblock \bibinfo{journal}{Il Nuovo Cimento} \bibinfo{volume}{12 C}
  (\bibinfo{year}{1989}).
\bibitem[{Abbasi et~al.(2011)}]{PhysRevD.83.012001}
\bibinfo{author}{R.~Abbasi}, et~al.,
\newblock \bibinfo{title}{{{Measurement of the atmospheric neutrino energy
  spectrum from 100~GeV to 400~TeV with IceCube}}},
\newblock \bibinfo{journal}{Phys.Rev.} \bibinfo{volume}{D83}
  (\bibinfo{year}{2011}) \bibinfo{pages}{012001}.
\bibitem[{Blobel(2002)}]{Blobel:2002p1184}
\bibinfo{author}{V.~Blobel},
\newblock \bibinfo{title}{{An Unfolding method for high-energy physics
  experiments}},
\newblock \bibinfo{journal}{{Contributions to the Conference on Advanced
  Statistical Techniques in Particle Physics, Durham}}  (\bibinfo{year}{2002})
  \bibinfo{pages}{258--267}.
\bibitem[{Muenich(2007)}]{Muenich:2007p1821}
\bibinfo{author}{K.~Muenich},
\newblock \bibinfo{title}{{Messung des atmosphaerischen Neutrinospektrums mit
  dem AMANDA-II Detektor}},
\newblock \bibinfo{journal}{Dissertation}  (\bibinfo{year}{2007}).
\bibitem[{{Adye}(2011)}]{Adye:2011p1825}
\bibinfo{author}{T.~{Adye}},
\newblock \bibinfo{title}{{{Unfolding algorithms and tests using RooUnfold}}},
\newblock \bibinfo{journal}{arXiv:1105.1160}  (\bibinfo{year}{2011}).

\end{thebibliography}
\end{document}